\def\mearth{{\rm\,M_\oplus}}

\def\msun{{\rm\,M_\odot}}

\def\gsim{~\rlap{$>$}{\lower 1.0ex\hbox{$\sim$}}}
\def\lsim{~\rlap{$<$}{\lower 1.0ex\hbox{$\sim$}}}

\def\eg{{\it e.g.\ }}

\def\ie{{\it i.e.\ }}
\def\cf{{\it c.f.\ }}

\documentclass{emulateapj}
\usepackage{graphicx}
\begin{document}

\title{Long-lived Chaotic Orbital Evolution of Exoplanets in Mean Motion Resonances with Mutual Inclinations}
\author{Rory Barnes\altaffilmark{1,2,3}, Russell Deitrick\altaffilmark{1,2}, Richard Greenberg\altaffilmark{4}, Thomas R. Quinn\altaffilmark{1,2}, Sean N. Raymond\altaffilmark{2,5}}
\altaffiltext{1}{Astronomy Department, University of Washington, Box 951580, Seattle, WA 98195}
\altaffiltext{2}{NASA Astrobiology Institute -- Virtual Planetary
Laboratory Lead Team, USA}
\altaffiltext{3}{E-mail: rory@astro.washington.edu}
\altaffiltext{4}{Lunar and Planetary Laboratory, 1629 E. University Blvd., Tucson, AZ 86716}
\altaffiltext{5}{CNRS, Laboratoire d'Astrophysique de Bordeaux, UMR 5804, F-33270, Floirac, France}

\begin{abstract}

We present N-body simulations of resonant planets with inclined orbits
that show chaotically evolving eccentricities and inclinations that
can persist for at least 10~Gyr. A wide range of behavior is possible,
from fast, low amplitude variations to systems in which eccentricities
reach 0.9999 and inclinations $179.9^\circ$. While the orbital
elements evolve chaotically, at least one resonant argument always
librates. We show that the HD~73526, HD~45364 and HD~60532 systems may
be in chaotically-evolving resonances. Chaotic evolution is apparent
in the 2:1, 3:1 and 3:2 resonances, and for planetary masses from
lunar- to Jupiter-mass. In some cases, orbital disruption occurs after
several Gyr, implying the mechanism is not rigorously stable, just
long-lived relative to the main sequence lifetimes of solar-type
stars. Planet-planet scattering appears to yield planets in inclined
resonances that evolve chaotically in about 0.5\% of cases. These
results suggest that 1) approximate methods for identifying unstable
orbital architectures may have limited applicability, 2) the observed
close-in exoplanets may be produced during the high eccentricity
phases induced by inclined resonances, 3) those exoplanets' orbital
planes may be misaligned with the host star's spin axis, 4) systems
with resonances may be systematically younger than those without, 5)
the distribution of period ratios of adjacent planets detected via
transit may be skewed due to inclined resonances, and 6) potentially
habitable planets in resonances may have dramatically different
climatic evolution than the Earth.  The {\it GAIA} spacecraft is
capable of discovering giant planets in these types of orbits.

\end{abstract}

\section{Introduction}
\label{sec:intro}

Exoplanetary systems with multiple planets show a wide variety of
orbital behavior, such as large amplitude oscillations of eccentricity
\cite[\eg][]{LaughlinAdams99,BG06}, mean motion resonances (MMRs)
\cite[\eg][]{LeePeale02,Raymond08}, and, in one case, oscillations of inclination
\citep{McArthur10,Barnes11}. Here we consider exoplanets in MMRs with mutual
inclinations and find that these systems can evolve with large,
chaotic fluctuations of eccentricity and inclination for at least 10
Gyr, but the MMR is maintained throughout.

In general, orbital interactions are broken into two main categories:
secular and resonant. The former treats the orbital evolution as
though the planets' masses have been spread out in the orbit, \ie the
mass distribution averaged over timescales much longer than the
orbital periods. If the orbits are non-circular or non-coplanar, the
gravitational forces change the orbital elements periodically.

MMRs occur when two or more orbital frequencies are close to integer
multiples of each other. The planets periodically reach the same
relative positions, introducing repetitive perturbations that can
dominate over secular effects. The combination drives the long-term
evolution of the system. For example, \cite{RiveraLissauer01}
considered the long-term behavior of Gl~876 \citep{Marcy01} and their
integrations show hints of chaotic evolution on 100~Myr timescales
(see their Fig.~4b). In this study we expand the research on exoplanets
in MMRs to include significant mutual inclinations.

While most bodies in our Solar System are on
low-$e$, low-$i$ orbits, the exoplanets do not share these
traits. Eccentricities span values from 0 to $>0.9$
\citep[\eg][]{Butler06}, and at least the two outer planets of
$\upsilon$ And have a large mutual inclination of $30^\circ$
\citep{McArthur10}. Numerous exoplanetary pairs in MMR are known,
including the systems of Gl~876, HD~82943, and HD~73526 in 2:1
\citep{Marcy01,Mayor04,Tinney06,Vogt05}, HD~45364 in 3:2
\citep{Correia09} and HD~60532 in 3:1 \citep{Desort08}. The orbital plane of Gl~876~b has been measured astrometrically with
{\it HST} \citep{Benedict02}, and it was reported that this
observation was compatible with a mutual inclination between orbital
planes
\citep{RiveraLissauer03}, but no formal publication demonstrated that
suggestion. The HD~128311 planetary system lies close to the 2:1
resonance, and recent astrometric measurements have measured the
orbital plane of HD~128311~c, but not the other planet
\citep{McArthur14}. Without knowledge of both orbital planes, the
mutual inclinations are unknown and use of the minimum masses and
coplanar orbits may not reveal the true orbital behavior, so this
system could in fact lie in the 2:1 resonance.

Several studies have also explored MMRs with mutual inclinations in
the context of planet formation. \cite{ThommesLissauer03} found that
convergent migration in a planetary disk can excite large inclinations
in exoplanetary systems. \cite{LeeThommes09} performed a
similar experiment and found that migrating planets could become
temporarily captured in an inclination-type MMR, in which conjunction librates about the midpoint between the longitudes of ascending node (inclination resonances are described in more detail in
$\S$~\ref{sec:methods}). \cite{LibertTsiganis09} explored the
formation of higher-order inclination resonances and also found that
temporary capture in an inclination resonance can occur. They also
artificially turned off migration shortly after capture and found the
resulting system was stable. \cite{TeyssandierTerquem14} considered
this process and identified several constraints on the planetary
mass ratio and eccentricities that permit entrance into an inclination
resonance. All these studies find that MMRs with inclination
can form in the protoplanetary disk, but only for a few
specific scenarios, and even then the inclination resonance is likely
to be fleeting. None of these studies considered the long-term
evolution of the resonant pairs.

Significant mutual inclinations may also be formed via gravitational
scattering events that typically result in the ejection of one planet
\citep{MarzariWeidenschilling02,Chatterjee08,Raymond10,Barnes11}. \cite{Raymond10}
also showed that scattering could produce systems in MMRs about 5\% of
the time, but they did not consider the inclinations of the resultant
systems. All these studies reveal that formation of MMRs with mutual
inclination is possible, but probably rare.

While {\it HST} has successfully characterized several nearby exoplanetary systems astrometrically, the {\it GAIA} space telescope could astrometrically detect hundreds
of Jupiter-sized planets \citep[see
\eg][]{Lattanzi00,Sozzetti01,Casertano08,Sozzetti14}, revealing how common are systems in an MMR with significant mutual inclinations. Against
this backdrop, we explore the dynamics of planets with orbital period
commensurabilities, significant eccentricities and mutual
inclinations.

This paper is organized as follows. In $\S$~\ref{sec:methods} we
describe the physics of resonance and our numerical methods. In
$\S$~\ref{sec:hypo}, we present results of hypothetical systems,
including planets in the habitable zone \citep{Kasting93,Kopparapu13},
as well as giant and dwarf planets in the 2:1, 3:1 and 3:2
commensurabilities. In $\S$~\ref{sec:form} we show that planet-planet
scattering can produce systems in the 2:1 MMR and with significant
mutual inclinations. In $\S$~\ref{sec:known} we analyze several known
exoplanetary systems and find several that could be evolving
chaotically. In $\S$~\ref{sec:discussion} we discuss the results and
then conclude in $\S$~\ref{sec:concl}.

\section{Methods}
\label{sec:methods}
\subsection{Resonant Dynamics}

MMRs are well-studied, and can be explained intuitively for low, but
non-zero, values of $e$ and inclination $i$
\citep{Peale76,Greenberg77,MurrayDermott99}. Stable resonances can be divided into two types: eccentricity
and inclination. The difference lies in the locations of the stable
longitudes of conjunction, sometimes called the libration center. In an eccentricity ($e$-type) resonance, the stable
longitudes are located at the longitude of periastron of the inner
planet ($\varpi_1$), and the apoastron of the outer planet ($\alpha_2
\equiv \varpi_2 + \pi$). For inclination ($i$-type) resonances, the
stable longitudes lie halfway between the longitudes of ascending
node, $\Omega$, of each planet ($\Omega_{1,2} \pm \pi/2$) when the
reference plane is the fundamental plane. When a system is formed, if
conjunction initially occurs near one of these stable points, and if
there is a commensurability of mean motions such that the conjunction
longitude varies slowly, then conjunction will tend to
librate. Furthermore, because the orbits are farthest apart at these
libration centers, resonances can reduce the likelihood of close
encounters, further maintaining long-term orbital stability.

Both $e$-type and $i$-type resonances are observed in our Solar
System, but $e$-type is far more prevalent, including the Galilean
satellite system and the Neptune-Pluto pair. An $i$-type resonance is
observed in the Saturnian satellite pair of Mimas and Tethys
\citep{Allan69,Greenberg73}. Neptune and Pluto are particularly relevant to this study, as the pair
is in the 3:2 $e$-resonance \citep{CohenHubbard65}, and later studies found that the $i$-resonance arguments also librate on short timescales, but the libration drifts slowly such that the resonant argument actually circulates with a
period of $\sim~25$~Myr \citep{WilliamsBenson71,Applegate86}. Higher order secular
resonances are also operating on Neptune and Pluto \citep{Milani89,KinoshitaNakai96}, and likely contribute to their configuration being formally chaotic
\citep{SussmanWisdom88}. Nonetheless, the pair is stable for at least
5.5 Gyr \citep{KinoshitaNakai96}. Thus, Pluto and Neptune's orbital
evolution is impacted by the $i$-resonance but it is likely a small
effect.

The first step in identifying a mean motion resonance is to examine
the ratio of orbital periods $P$ or, equivalently, the mean motions $n
\equiv 2\pi/P$. If the ratio is close to a ratio of two integers, \ie
$n_1/n_2 \approx j_1/j_2$ where $j_1$ and $j_2$ are integers, then the system
{\it may} be in resonance. An MMR requires a periodic force to applied
near the same longitude relative to an apse or a node, which precess
with time. To account for this evolution, celestial mechanicians have
introduced the resonant argument, a combination of angles that account
for both the mean motions and the orbital orientations.

For $e$-type resonances, the resonant arguments are
\begin{equation}
\theta_{1,2} = j_1\lambda_2 - j_2\lambda_1 - j_3\varpi_{1,2},
\label{eq:e_res_arg}
\end{equation}
where the $j$ terms are integers that sum to 0, and the subscripts 1
and 2 to $\varpi$ correspond to the inner and outer planet,
respectively. Should any $\theta$ librate with time, or even circulate
slowly, then resonant dynamics are important. Conjunction ($j_1\lambda_2
- j_2\lambda_1$) lies close to a particular longitude relative to the
apsides. Usually $\theta$ will librate about either 0 or $\pi$, but
other stable values are possible and are referred to as asymmetric
resonances.

For $i$-type resonances, the dominant resonant argument is
\begin{equation}
\phi = j_1\lambda_2 - j_2\lambda_1 - j_3\Omega_1 - j_4\Omega_2,
\label{eq:i_res}
\end{equation}
if $i=0$ corresponds to the
invariable, or fundamental, plane, \ie the plane perpendicular to the
total angular momentum of the system. With this choice, $\Omega_1 =
\Omega_2~\pm~\pi$. Inclination resonances are fundamentally
different from $e$-type in that first order resonances are technically
not possible, \ie $j_1-j_2 > 1$. The stable conjunction longitudes
correspond to the two mid-node longitudes. Note that for circular orbits, there is no
substantive difference between the two stable longitudes: The geometry at $\pm~90^\circ$
from the mutual node are identical.  See \cite{Greenberg77} for more discussion on
the physics of the $i$-resonance.

It is often convenient to think of resonances in terms of a
pseudo-potential. The conjunction longitude, $\lambda_c$ is
accelerated toward the stable points, and oscillates sinusoidally
about it. The acceleration of $\lambda_c$ in a 2:1
(4:2) MMR can be written as
\begin{equation}
\ddot{\lambda_c} = c_1e_1\sin(\lambda_c - \varpi_1) + c_2e_2\sin(\varpi_2 - \lambda_c) + c_3i_1^2\sin 2(\lambda_c - \Omega_1),
\label{eq:potential}
\end{equation}
where $c_1, c_2$ and $c_3$ are constants that depend on the masses and semi-major axes \citep{Greenberg77}, and $\lambda_c = 2\lambda_2 - \lambda_1$. Eq.~(\ref{eq:potential}) is analogous to that of a compound pendulum, as long
as the $e$'s and $i$'s are approximately constant. The
pseudo-potential contains minima at $\varpi_1$, $\alpha_2$, and
$\Omega_1~\pm~\pi/2$, but each has different depths which vary as the
orbits evolve. Moreover, as $e$ and/or $i$ become large, this simple
picture breaks down and more minima will likely appear. Thus, we should anticipate
the motion to become complicated, an expectation that is borne out in
$\S\S$ \ref{sec:hypo} -- \ref{sec:known}.

\subsection{Numerical Methods}

The classical analytic theory summarized above becomes less accurate as the values of eccentricity and inclination increase. To lowest order, the
evolution of $e$ and $i$ are decoupled, but as either or both become
large, pathways for the exchange of angular momentum open up, and the
motion can become very complicated \citep[\eg][]{Barnes11}. Thus, we rely on $N$-body numerical methods to analyze
resonances with mutual inclinations, while appealing to published analytic theory to help interpret the outcomes. In particular we use the well-tested and reliable
\texttt{Mercury} \citep{Chambers99} and \texttt{HNBody}
\citep{RauchHamilton02} codes. We do not include general relativistic corrections, which are mostly negligible for the planetary systems we consider here. We used both mixed variable symplectic and
Bulirsch-Stoer methods to integrate our systems. While the former can
integrate our hypothetical systems to 10 Gyr within about 10 days on a
modern workstation, the Bulirsch-Stoer method, which is more accurate,
requires about 80--90 days with \texttt{HNBody}; hence we used it
sparingly. Regardless of our choice of software and/or integration
scheme, identical initial conditions produced qualitatively similar
results. All integrations presented here conserved energy to better
than 1 part in $10^5$, usually better by many orders of
magnitude. Unless stated otherwise, all systems maintained an MMR for
10 Gyr and met our energy conservation requirements. The reference
plane for $i$ and $\Omega$ for all cases is the invariable plane so
that the inclinations and longitudes of ascending node are more
physically meaningful\footnote{We have made our source code to rotate
any astrocentic orbital elements into the invariable plane, as well as
generate input files for \texttt{Mercury} and
\texttt{HNBody}, publicly available at https://github.com/RoryBarnes/InvPlane.}. Here we ignore planetary and stellar spins: Planets and stars are point masses.

We consider several broad categories of exoplanetary systems. First we
model systems in the 2:1 resonance. In Set \#1, one planet is always 1
Earth-mass ($1~\mearth$) with a semi-major axis $a$ of 1 AU, and the
primary is a solar-mass star, \ie it is in the habitable zone
\citep{Kasting93,Kopparapu13}. The other planet may have a mass
between 3 and 40~$\mearth$. In Set \#2, we explore the case of two giant
planets orbiting a 0.75~$\msun$ star in 2 and 4 year orbits. Set \#3
considers two dwarf planets orbiting a solar-mass star.  In Sets \#4
and \#5, we simulated $\sim$Earth-mass planets in the 3:1 and 3:2 MMR,
respectively. For all these simulations, the period ratios are at
exact resonance, but the other orbital elements are chosen randomly
and uniformly over a wide range of values. For each set, we integrated
100 systems for 10~Myr as an initial survey, and for systems that
appeared interesting, \ie showed chaotic evolution, we continued them
to 10 Gyr using 0.01~year timesteps. Table~\ref{tab:hypo} shows the ranges of
initial conditions we used for these sets of hypothetical systems.

\begin{deluxetable*}{cccccccccccc}
\tabletypesize{\footnotesize}
\tablecolumns{12}
\tablecaption{Initial Conditions for Hypothetical Systems}
\tablehead{\colhead{Set} & \colhead{MMR} & 
	\colhead{$M_*$ (M$_\odot$)} & \colhead{$m_1$} & \colhead{$m_2$} & 
	\colhead{$a_1$ (AU)} & \colhead{$a_2$ (AU)} & \colhead{$e$} & 
	\colhead{$i$ ($^\circ$)} & \colhead{$\Omega$ ($^\circ$)} & 
	\colhead{$\omega$ ($^\circ$)} & \colhead{$\lambda$ ($^\circ$)}}
\startdata
1 & 2:1 & 1 & 1 M$_\oplus$ & 3.3--36.6 M$_\oplus$ & 1 & 1.5874 & 0--0.5 & 0--30 & 0--360 & 0--360 & 0--360\\
2 & 2:1 & 0.75 & 0.314 M$_{Jup}$ & 0.1--1.15~M$_{Jup}$ & 1.44219 & 2.2893 & 0--0.5 & 0--30 & 0--360 & 0--360 & 0--360\\
3 & 2:1 & 1 & 0.0775 M$_{Moon}$ & 2.84 M$_{Moon}$ & 1 & 1.5874 & 0--0.5 & 0--30 & 0--360 & 0--360 & 0--360\\
4 & 3:1 & 1 & 1 M$_\oplus$ & 3.3--36.6 M$_\oplus$ & 1 & 2.0801 & 0--0.5 & 0--30 & 0--360 & 0--360 & 0--360\\
5 & 3:2 & 1 & 1 M$_\oplus$ & 3.3--36.6 M$_\oplus$ & 1 & 1.3104 & 0--0.5 & 0--30 & 0--360 & 0--360 & 0--360\\
\label{tab:hypo}
\end{deluxetable*}

In $\S$~5 we examine known systems in or near the 2:1 (HD~73526 and
HD~128311), 3:2 (HD~60532) and 3:1 (HD45364) MMRs. Each of these
systems was discovered via radial velocity data, and hence suffer from
the mass-inclination degeneracy. However, HD~128311~c has also been
detected astrometrically with {\it HST} \citep{McArthur14}, hence its
mass and full orbit are known. The best fits and uncertainties in parentheses for
these known systems are listed in Table~\ref{tab:known}. The position of each planet
in its orbit, the ``phase,'' is crucial information and different
authors present the information in different formats and so we present
the parameter used in the most recent paper in Table 2. $T_p$ is the
time of periastron passage, $\lambda$ is the mean longitude and $\mu$
is the mean anomaly.

For each of the known systems, we vary each orbital element uniformly
within its published uncertainties and simulate the orbital evolution
with \texttt{HNBody}. Our goal is not to calculate a probability that
each system is in an $i$-type resonance, but rather to determine if it
is possible at all. We simulate 100 versions of the these systems and allow $i$ and $\Omega$ to take any value, and $m$ is adjusted
accordingly. If chaotic motion in the MMR is apparent, we integrate it
for a further 10 Gyr.

\begin{deluxetable*}{cccccccc}
\tabletypesize{\footnotesize}
\tablecolumns{8}
\tablecaption{Best Fits and Uncertainties for Selected Known Exoplanet Systems} 

\tablehead{\colhead{System} & \colhead{$M_*$ (M$_\odot$)} & 
	\colhead{Planet} & \colhead{$m$ (M$_{Jup}$)} & \colhead{$P$ (d)} & 
	\colhead{$e$} & \colhead{$\omega$ ($^\circ$)} & \colhead{Phase}}
\startdata
HD~128311 & 0.828 & b & 1.769 (0.023) & 453.019 (0.404) & 0.303 (0.011) & 57.864 (3.258) & 2400453.019 (4.472)$^a$\\
 & & c & 3.125 (0.069) & 921.538 (1.15) & 0.159 (0.006) & 15.445 (6.87) & 2400921.538 (18.01)$^a$\\
HD~73526 & 1.08 & b & 2.8 (0.2) & 188.3 (0.9) & 0.19 (0.05) & 203 (9) & 86 (13)$^b$\\ 
 & & c & 2.5 (0.3) & 377.8 (2.4) & 0.14 (0.09) & 13 (76) & 82 (27)$^b$\\
HD~60532 & 1.44 & b & 1.03 (0.05) & 201.3 (0.6) & 0.28 (0.03) & 351.9 (4.9) & 2453987 (2)$^a$\\
 & & c & 2.46 (0.09) & 604 (9) & 0.02 (0.02) & 151 (92) & 2453723 (158)$^a$\\
HD~45364 & 0.82 & b & 0.1872 (0) & 226.93 (0.37) & 0.1684 (0.019) & 162.58 (6.34) & 105.76 (1.41)$^c$\\
 & & c & 0.6579 (0) & 342.85 (0.28) & 0.0974 (0.012) & 7.41 (3.4) & 269.52 (0.58)$^c$\\
\tablenotetext{a}{$T_{peri}$ (JD)}
\tablenotetext{b}{$\mu$ ($^\circ$)}
\tablenotetext{c}{$\lambda$ ($^\circ$)}
\label{tab:known}
\end{deluxetable*}

\begin{deluxetable*}{ccccccccc}
\tabletypesize{\normalsize}
\tablecaption{Initial Conditions for Selected Set \#1 Systems}
\tablecolumns{9}
\tablehead{\colhead{System} & \colhead{Body} & \colhead{$m$ ($M_\oplus$)} & 
	\colhead{$a$ (AU)} & \colhead{$e$} & \colhead{$i$ ($^\circ$)} & 
	\colhead{$\Omega$ ($^\circ$)} & \colhead{$\omega$ ($^\circ$)} & 
	\colhead{$\mu$ ($^\circ$)}}
\startdata
A & 1 & 1 & 1 & 0.0866 & 10.322 & 223.75 & 340.2 & 268.33\\        
  & 2 & 10.07 & 1.5874 & 0.1883 & 0.82 & 43.747 & 74.21 & 296.5\\
B & 1 & 1 & 1 & 0.0251 & 1.61 & 277.19 & 353.67 & 261.33\\         
  & 2 & 26.91 & 1.5874 & 0.0531 & 0.0475 & 97.19 & 219.07 & 359.38\\
C & 1 & 1 & 1 & 0.2373 & 2.918 & 217.13 & 330.25 & 102.45\\        
  & 2 & 4.39 & 1.5874 & 0.4225 & 0.565 & 37.13 & 246.36 & 112.88\\
D$^b$ & 1 & 1 & 1 & 0.00296 & 19.83 & 278.01 & 332.27 & 268.52\\   
  & 2 & 35.62 & 1.5874 & 0.266 & 0.449 & 98.01 & 69.73 & 343.09\\ 
E & 1 & 1 & 1 & 0.2952 & 38.51 & 247.92 & 40.44 & 225.93\\         
  & 2 & 14.82 & 1.5874 & 0.0961 & 1.832 & 67.92 & 253.92 & 318.12\\
F & 1 & 1 & 1 & 0.1 & 12.04 & 270.26 & 140.77 & 0\\                
  & 2 & 10 & 1.5874 & 0.15 & 0.9551 & 90.26 & 23.77 & 10\\
G & 1 & 1 & 1 & 0.15 & 23.526 & 232.96 & 31 & 0\\                  
  & 2 & 10 & 1.5874 & 0.2 & 1.832 & 52.96 & 247.9 & 10\\
\tablenotetext{b}{Stable for only 73~Myr.}
\label{tab:set1}
\end{deluxetable*}

\section{Hypothetical Systems}
\label{sec:hypo}

In this section we present the results of the simulations described in
the previous section. We separate the results by resonance: 2:1, then 3:1,
and finally 3:2. In all cases we find evidence of chaotic orbital
evolution, often with very large amplitudes of eccentricity and
inclination.

\subsection{The 2:1 Resonance} 

In this section we examine example cases in Sets \#1--3, \ie in the 2:1 MMR. The $e$-resonance arguments are 
\begin{equation}
\theta_{1,2} = 2\lambda_2 - \lambda_1 - \varpi_{1,2},
\label{eq:e_res_2:1}
\end{equation}
and the $i$-resonance argument is
\begin{equation}
\phi = 4\lambda_2 - 2\lambda_1 - \Omega_1 - \Omega_2.
\label{eq:i_res_2:1}
\end{equation}

\subsubsection{Earth with an Exterior Companion (Set \#1)}

\begin{figure*}
\includegraphics[width=0.99\textwidth]{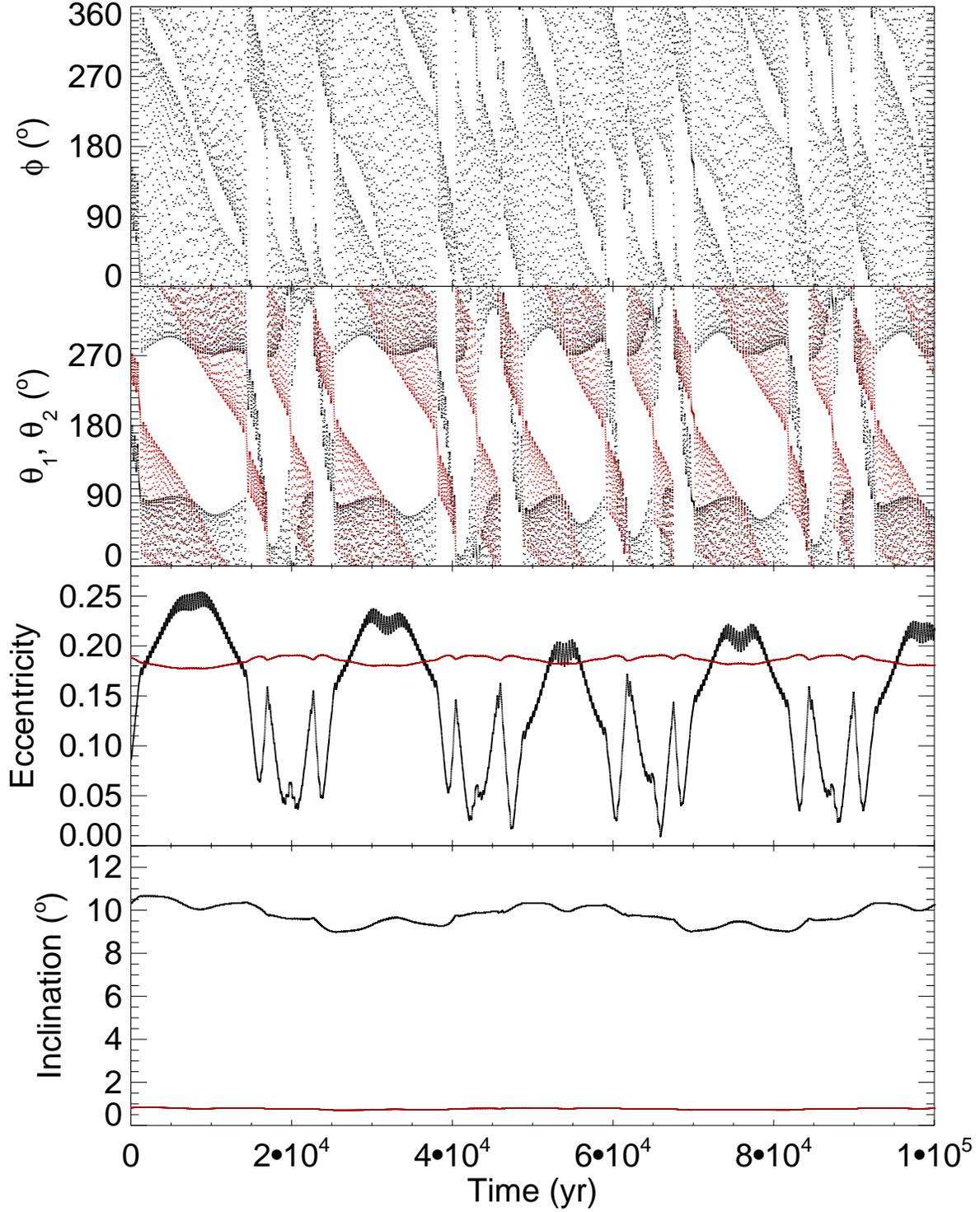}
\caption{The first $10^5$ years if evolution of System A. Black corresponds to the inner Earth-mass planet initially at 1 AU, red to a larger planet in the outer 2:1 resonance. Variations of the inclination resonance argument (top), eccentricity arguments (top middle) with black dots corresponding to the $\varpi_1$ argument and red to $\varpi_2$, eccentricities (bottom middle) and inclinations (bottom).}
\label{fig:105_100kyr}
\end{figure*}

Set \#1 consists of an Earth-mass planet with a 1 year period and
a larger exterior planet with an orbital period of 2 years. In
Fig.~\ref{fig:105_100kyr} we show the evolution of the resonant
arguments, $e$ and $i$ for the first example, System A in Table~\ref{tab:set1}, for
the first $10^5$ years.

The top two panels show that the resonance arguments switch between
libration and circulation. Initially $\theta_1$ (black dots) librates
about 0, the classic libration center for $e$-type resonances, while
$\theta_2$ (red dots) librates on short timescales, but circulates on
long timescales. Also note that the $e$- and $i$-resonance arguments
appear to be coupled. During this initial phase, $\phi$ librates with
large amplitude.  After about 14,000 years, the behavior changes
dramatically and all arguments librate, but with sudden jumps between
libration centers. Then at 25,000 years, the motion appears to return
to the initial state. This switching between modes of oscillation
demonstrates the system is chaotic, and is often an indicator of
impending instability \citep[\eg][]{Laskar90}. Hence we would naively
expect a similar outcome for this system.  The eccentricity of the
inner planet shows unusual behavior that also changes with the
resonance arguments. The evolution is approximately periodic, but does
not appear regular. The inclinations do not appear to be strongly
impacted by the changes in the resonance argument behavior.

\begin{figure*}
\includegraphics[width=0.99\textwidth]{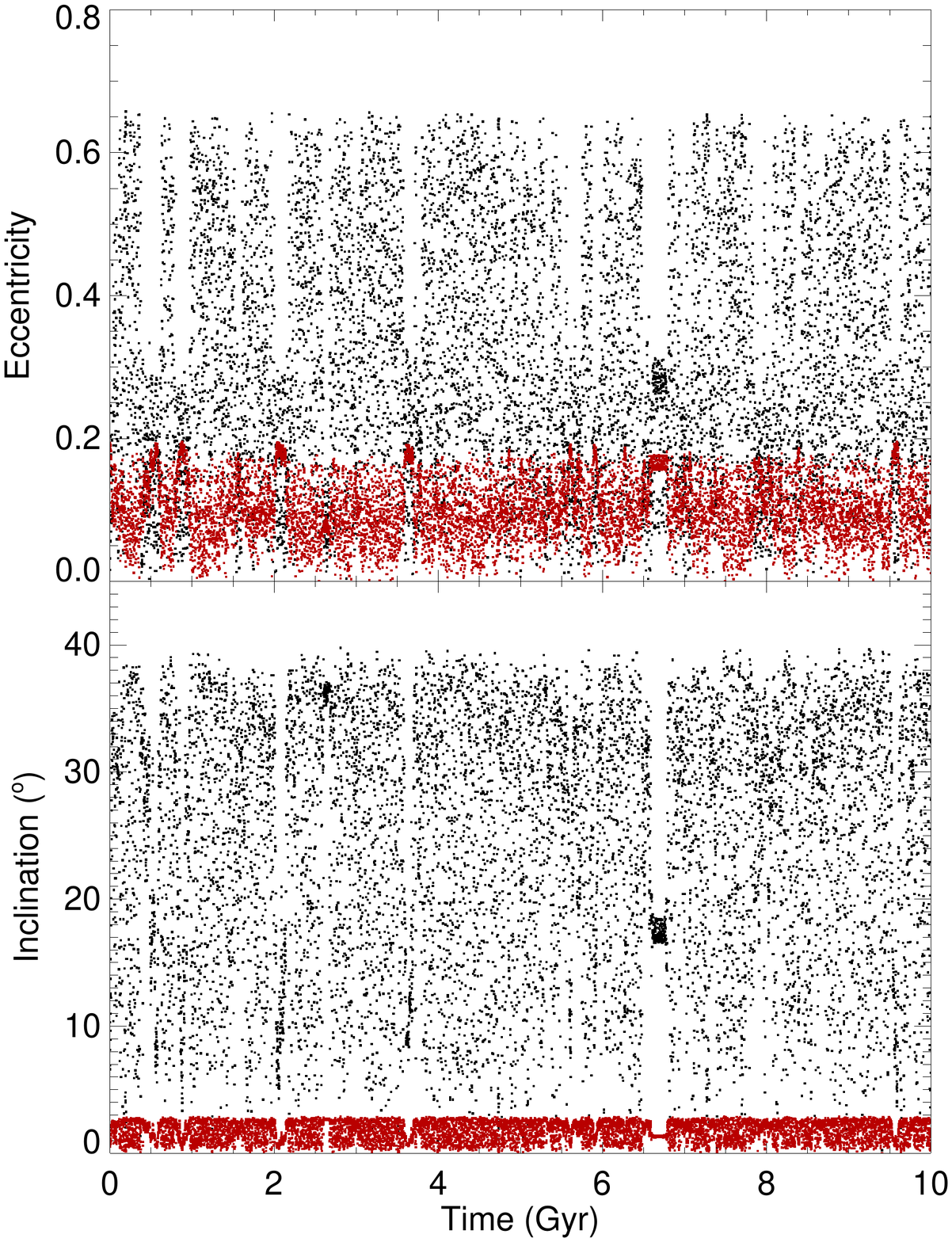}
\caption{The evolution of $e$ and $i$ for System A over 10 Gyr. Black points correspond to the inner planet and red to the outer.}
\label{fig:105_10Gyr}
\end{figure*}

In Fig.~\ref{fig:105_10Gyr}, we extend the evolution of $e$ and $i$ by
a factor of $10^5$, to 10~Gyr. Here we see that the hints of chaotic
behavior in Fig.~\ref{fig:105_100kyr} remain present, but are not
indicative of the true scale of the chaotic motion for this
system. Remarkably, the system survives for 10~Gyr despite the
chaos. The eccentricity of the inner planet aperiodically reaches
values larger than 0.65, while its inclination reaches $40^\circ$. The
outer planet is more massive, so its evolution does not have as large
an amplitude as the inner, but it, too, evolves chaotically. While the
variations in $e$ and $i$ are chaotic, there are clearly bounds to
their permitted values.

\begin{figure*}
\begin{tabular}{cc}
\includegraphics[width=0.49\textwidth]{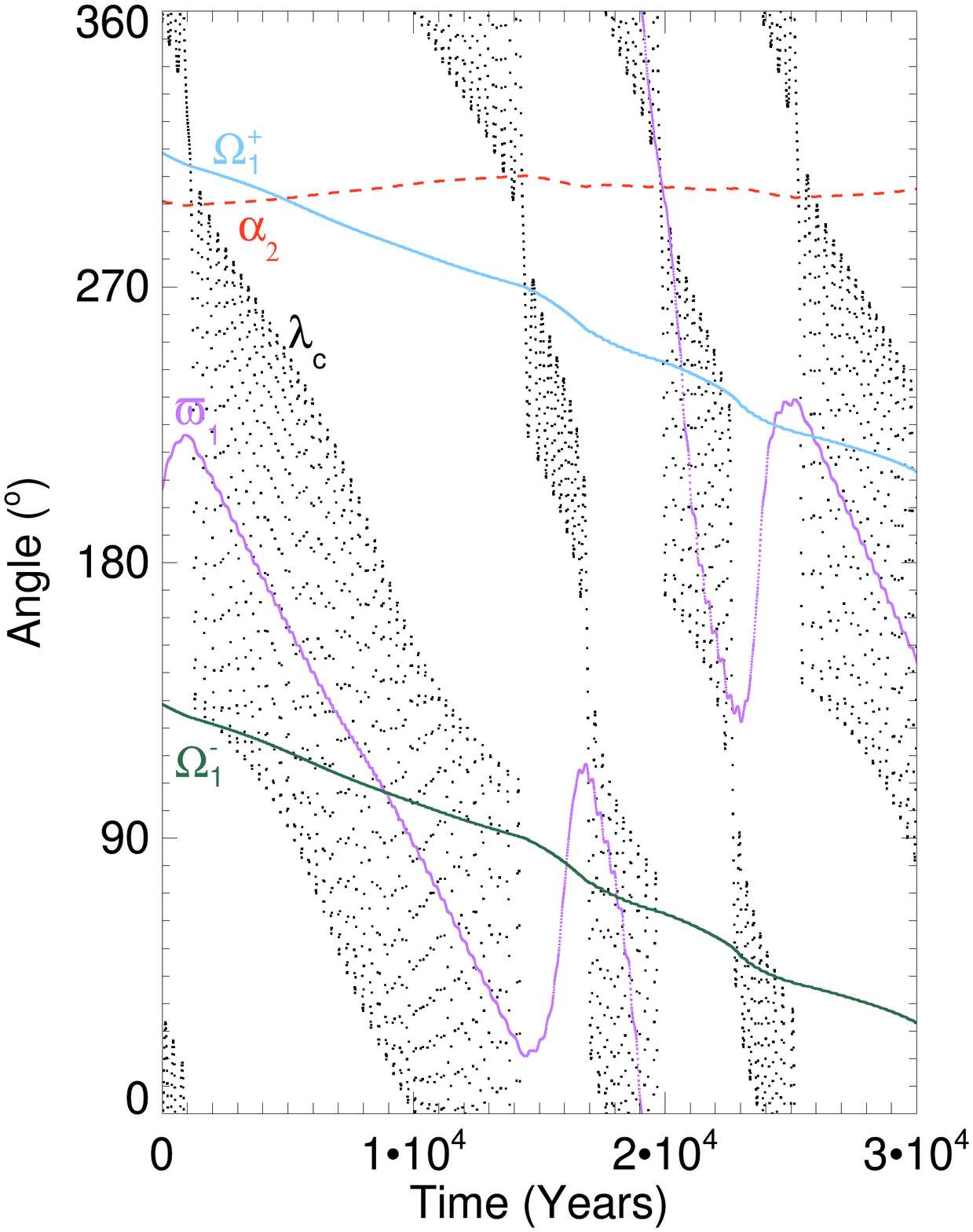} &
\includegraphics[width=0.49\textwidth]{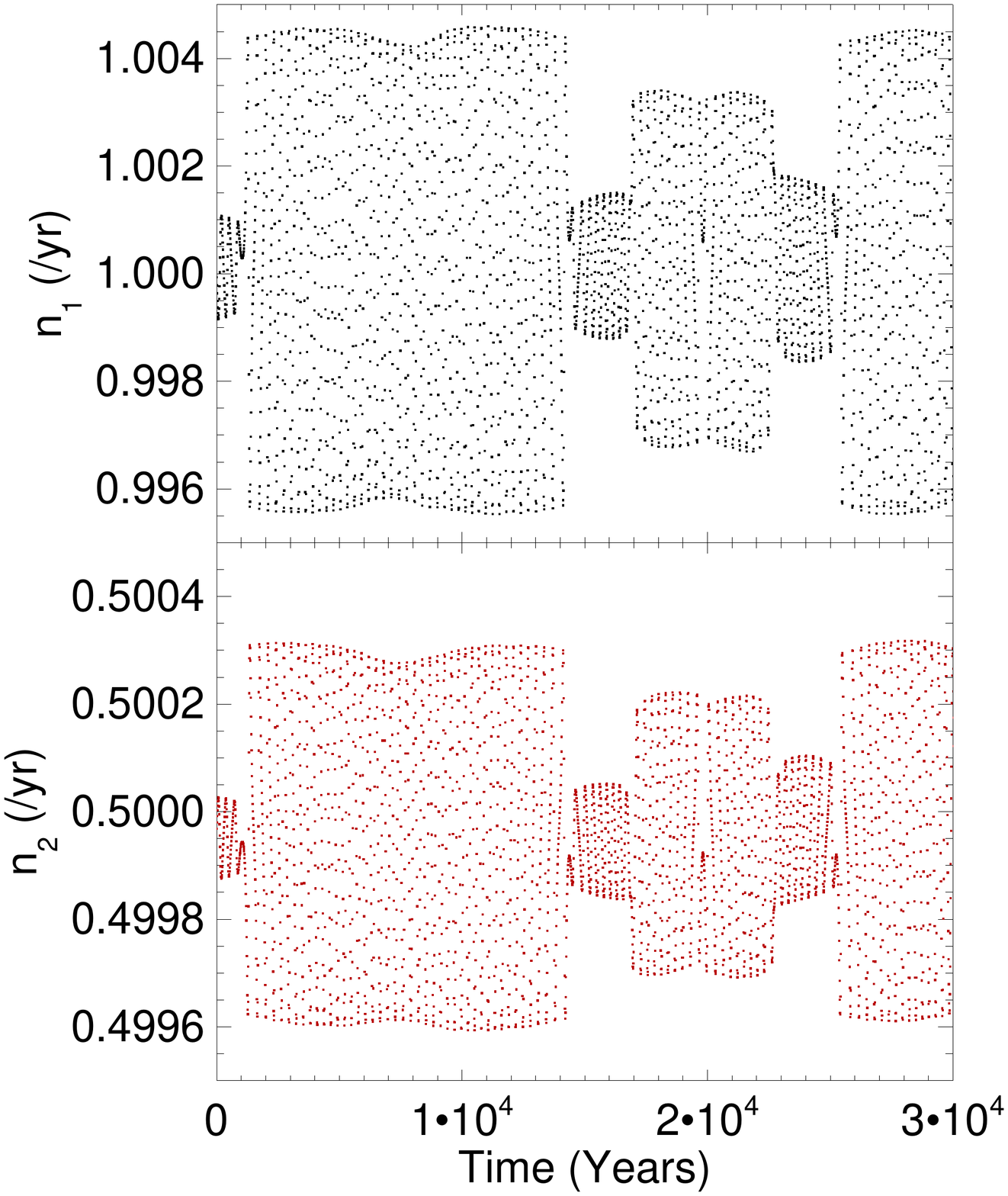} 
\end{tabular}
\caption{Initial evolution of System A. {\it Left:} Evolution of the conjunction longitude $\lambda_c$ (black dots) in relation to the 4 classic stable longitudes, $\varpi_1$ (purple curve),
$\alpha_2~\equiv~\varpi_2+\pi$ (dashed orange), $\Omega_1^+~\equiv~\Omega_1~+~\pi/2$
(blue), and $\Omega_1^-~\equiv~\Omega_1~-~\pi/2$ (green). {\it Right:} Evolution of the mean motions of the inner planet (top) and outer planet (bottom).}
\label{fig:indepth} 
\end{figure*}

The libration and slow circulation of the resonant arguments suggest
that both $e$ and $i$-type resonances are important in this system. We
further explore the evolution of the system in
Fig.~\ref{fig:indepth}, which only considers the first 30,000
years of orbital evolution. In the left panel we compare the
conjunction longitude (black dots) with the various stable longitude
predicted from classic resonant theory, $\varpi_1$ (purple curve),
$\alpha_2~\equiv~\varpi_2+\pi$ (orange curve), $\Omega_1^+~\equiv~\Omega_1~+~\pi/2$
(blue curve), and $\Omega_1^-~\equiv~\Omega_1~-~\pi/2$ (green curve). Conjunction librates, but, surprisingly, not always about the expected stable
longitudes. For most of the time, $\lambda_c~\sim~\varpi_1$, but it
also librates about other locations that are not associated with any
classic libration center. Note that conjunction avoids $\alpha_2$.

In the right panel, we plot the two mean motions and see that they
librate about the resonant frequencies, but with varying
amplitudes. Careful inspection of the two panels shows that the
different mean motion amplitudes correspond to the different
librations seen in the left panel. There appear to be 5 different
resonant oscillations over this cycle, which approximately repeats
for at least the first 1 Myr.

\begin{figure*}
\includegraphics[width=0.99\textwidth]{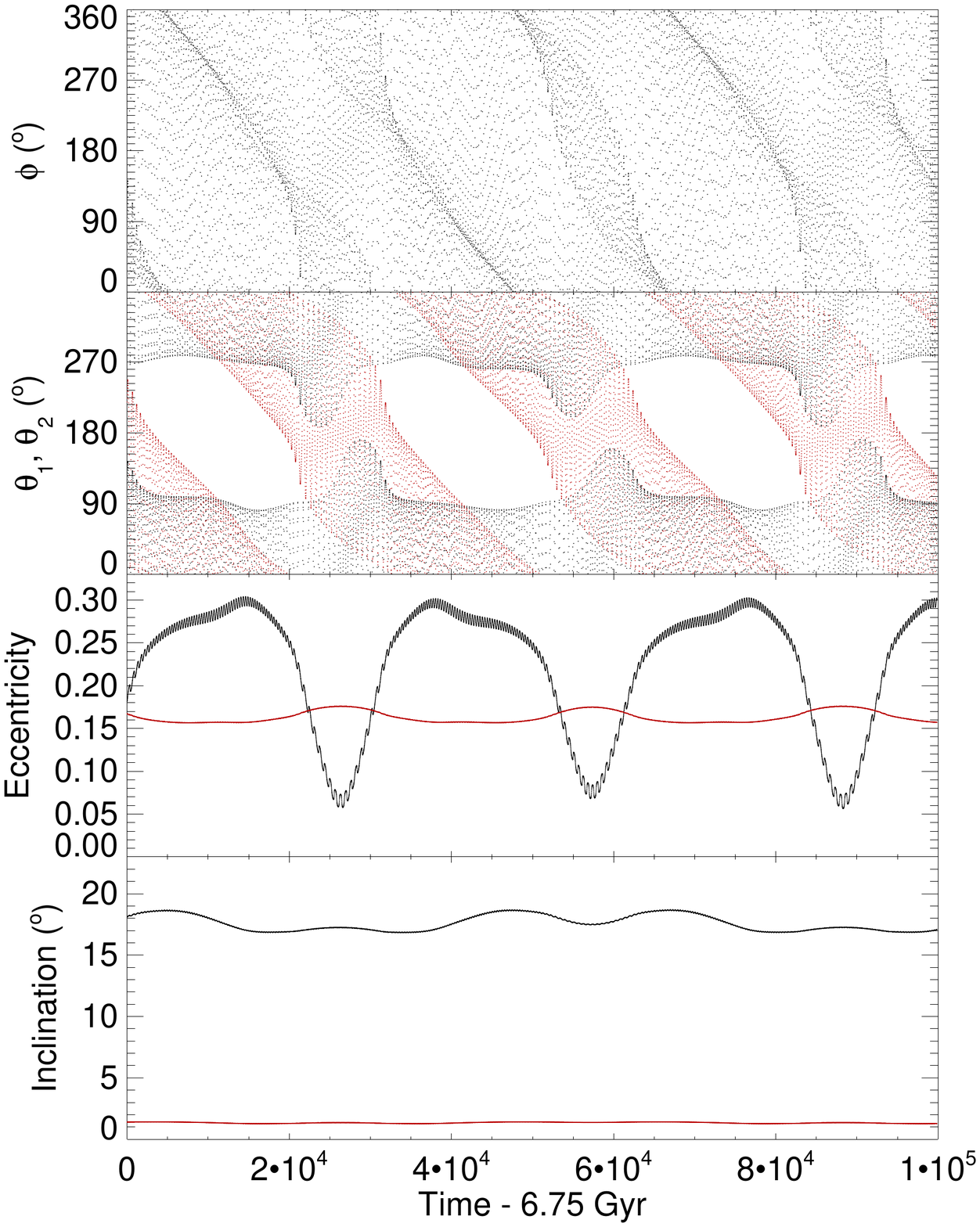}
\caption{Same as Fig.~\ref{fig:105_100kyr}, but at 6.75~Gyr.}
\label{fig:105_6.75Gyr}
\end{figure*}

The long-term behavior in Fig.~\ref{fig:105_10Gyr} shows mode-switching can
occur on longer timescales as well. For example, from 6.5 -- 6.8 Gyr,
$e$ and $i$ are confined to narrow regions, but then return to the
large amplitude oscillations. Fig.~\ref{fig:105_6.75Gyr} shows the
evolution over $10^5$ years starting at 6.75~Gyr within this
alternative mode. In comparison to 
Fig.~\ref{fig:105_100kyr}, the $i$-resonance argument is circulating, as is
$\theta_2$. However, $\theta_1$ is librating about 0 indicating that
the system is exclusively in an $e$-type resonance.

After 10~Gyr the evolution is qualitatively similar to the initial
evolution. The $e$- and $i$-resonant arguments are switching between
libration and circulation, a quasi-stable behavior. We are unaware of
any study that has found qualitatively similar behavior in a planetary
or satellite system. While long-lived chaos is evident in our Solar
System \citep[\eg][]{SussmanWisdom88}, the amplitudes of the
variations are much smaller. Moreover, the switching between different
quasi-periodic states is also usually an indicator of
instability. System A is not ``nearly integrable,'' as is often argued
for stable planetary systems. Nonetheless, these planets remains bound
and in resonance for the main sequence lifetime of a solar-type star.

\begin{figure*} 
\includegraphics[width=0.99\textwidth]{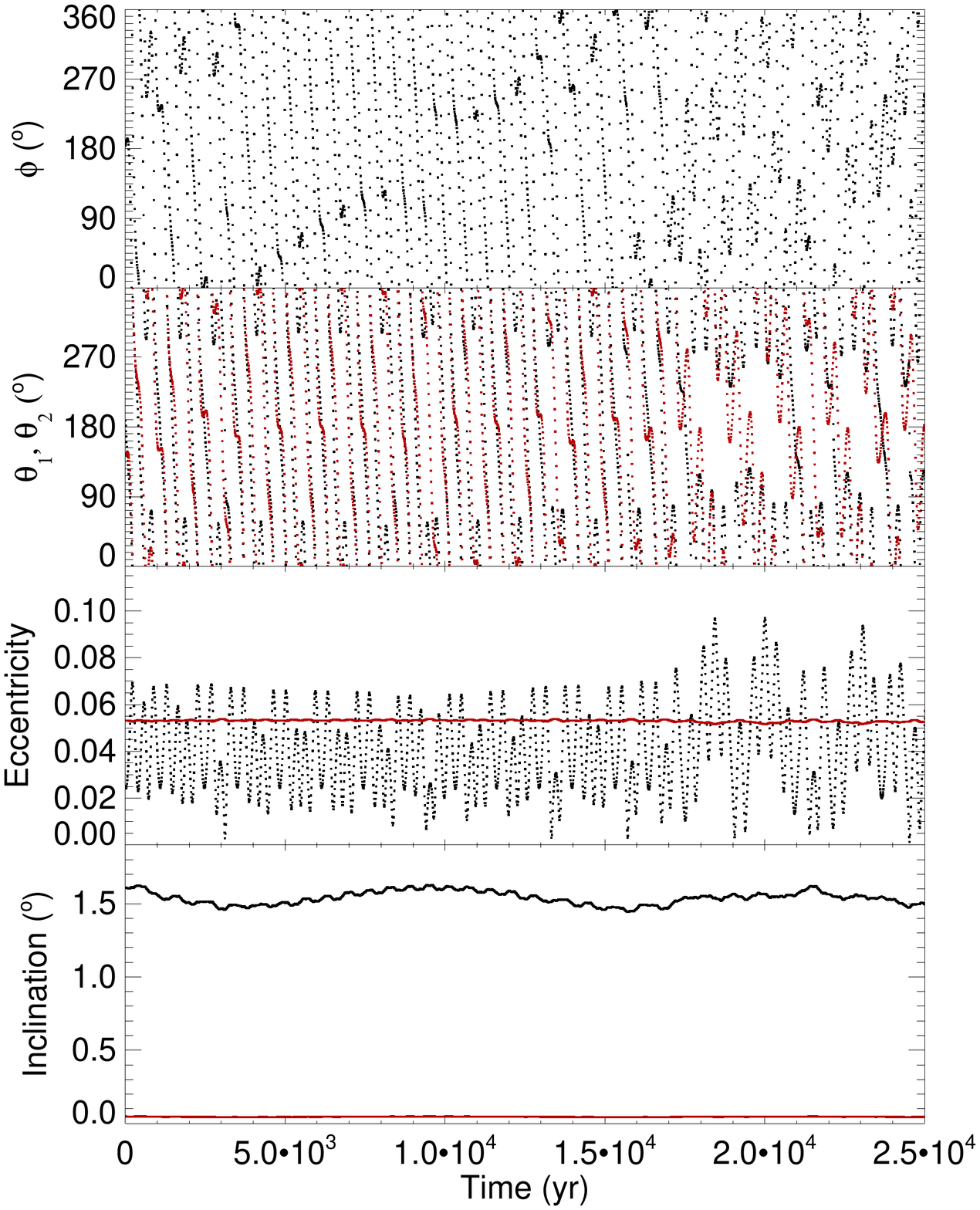}
\caption{The first 25 kyr of evolution of System B in the same format as Fig.~\ref{fig:105_100kyr}.}
\label{fig:104}
\end{figure*}

\begin{figure*}
\includegraphics[width=0.99\textwidth]{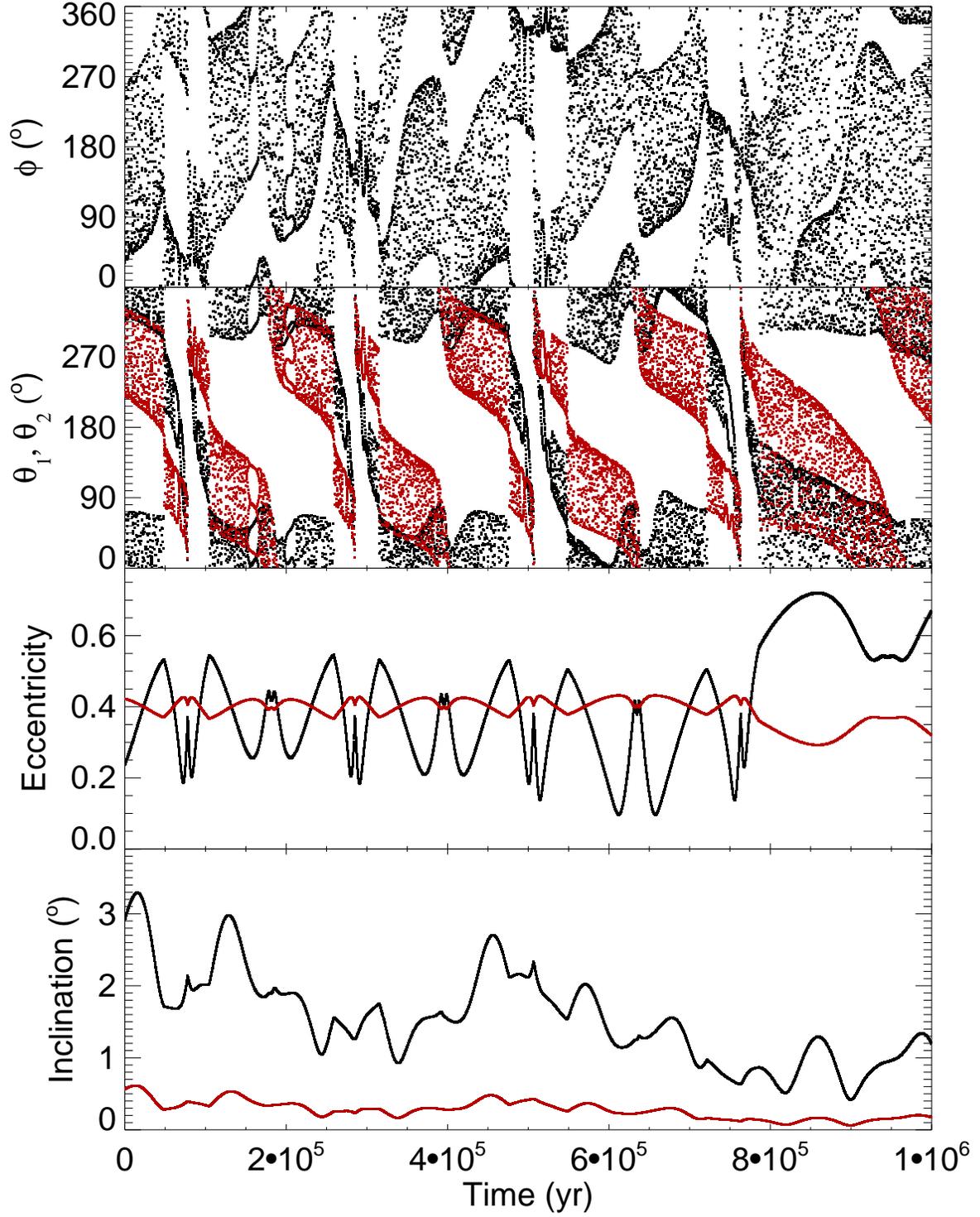}
\caption{The first 1~Myr of evolution of System C in the same format as Fig.~\ref{fig:105_100kyr}.} 
\label{fig:120}
\end{figure*}

\begin{figure*}
\includegraphics[width=0.99\textwidth]{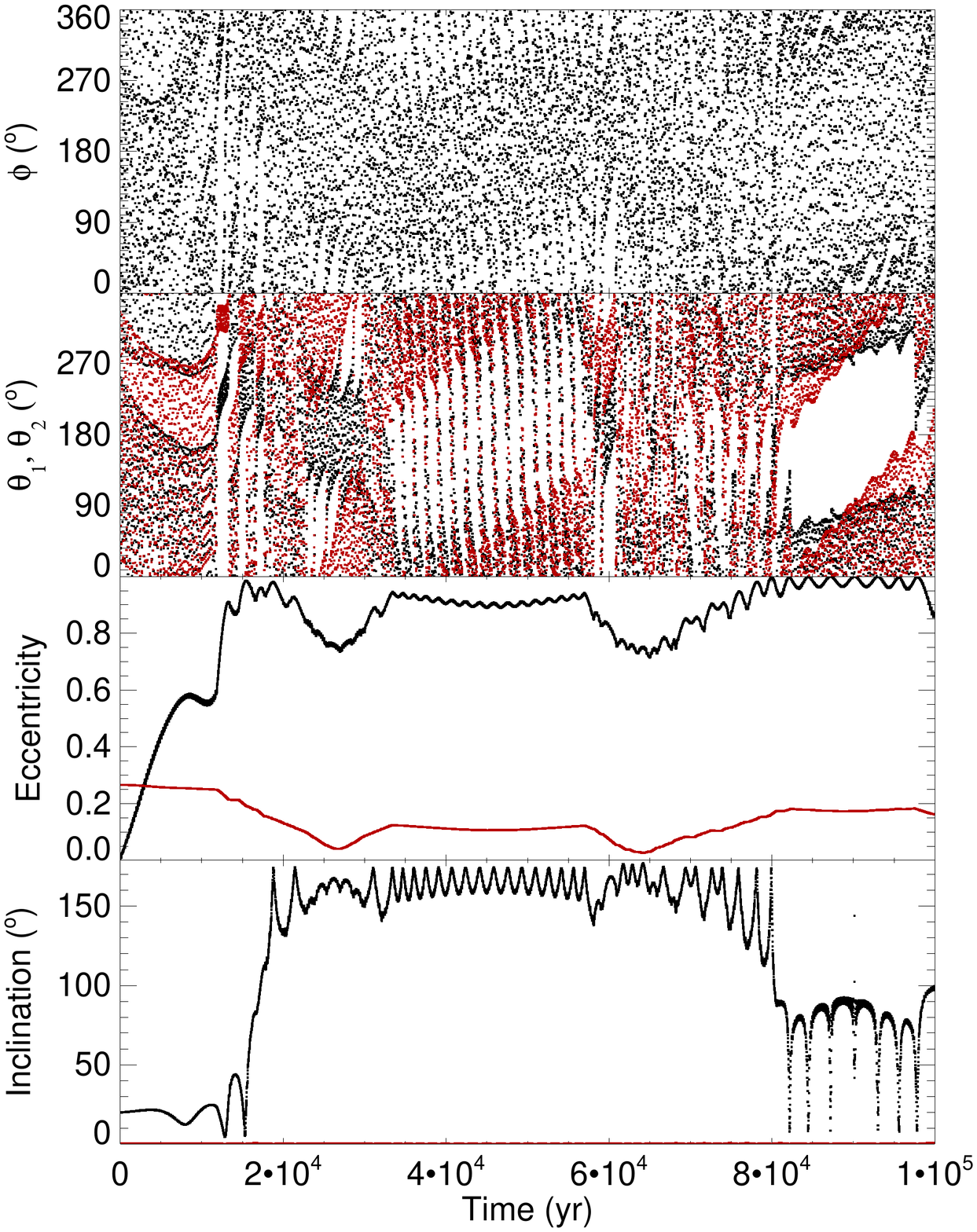}
\caption{The first $10^5$ years of evolution of System D in the same format as Fig.~\ref{fig:105_100kyr}.  This system destabilized after 73~Myr.}
\label{fig:109}
\end{figure*}

Although the orbital behavior of System A is surprising, this case is
not anecdotal. Table~\ref{tab:set1} contains 6 examples from Set \#1 that show
chaotic evolution, yet remain in resonance for 10~Gyr. In
Figs.~\ref{fig:104}--\ref{fig:109}, we show 3 more cases as
representative examples of the 2:1 MMR. These simulations demonstrate that chaotic inclined MMRs can produce behavior that spans a range
from high frequency, low amplitude oscillations to low frequency and high amplitude oscillations to cases in which all available phase space is sampled.

System B is an example of high frequency, low amplitude
evolution. Fig.~\ref{fig:104} shows the first 25,000 years of its
evolution. In this case the resonant arguments oscillate with a period
of a few hundred years and switch states at $\sim$18,000 years. The
eccentricities and inclinations of the Earth-like planet oscillate
with an amplitude of 0.1 and 0.25$^\circ$, respectively, on this
timescale. Note that System B begins with a mutual inclination of just
1.65$^\circ$, revealing that relatively small mutual inclinations can
lead to chaotic orbital evolution. Over 10 Gyr, $e_1$ remains below 0.14 and $e_2$ 0.06, while $i_1$ remains below $18^\circ$ and $i_2$ below $1^\circ$.

System C is similar to System A in its evolution, as shown in
Fig.~\ref{fig:109}. Over 10 Gyr $e_1$ varies from 0 to 0.9 and $i_1$
varies from 0 to $60^\circ$. From $\sim 3$~Gyr to $\sim 6$~Gyr (not shown) the
system enters a different mode in which the eccentricities and
inclinations are confined to narrower regions.

System D only survives in the resonance for 73~Myr, but we it include
here to illustrate the extreme eccentricity evolution that is possible
in inclined MMRs. At 87,170 years, $e_1$ reaches a value of 0.99998,
implying a periastron distance that would place it inside the {\it
core} of its solar-mass primary. Clearly, the point-mass approximation
for the orbital dynamics has broken down for this system. In
particular, we expect that at when $e_1~\sim~1$ that strong tidal
effects should dramatically alter the orbits. We return to this point in $\S$~\ref{sec:discussion}.

Rather than show similar plots for Systems E--G, we only comment on
their behavior. System E shows a circulating $\phi$ for the first
$10^5$ years, while $\theta_1$ librates with large amplitude about 0,
and $\theta_2$ drifts. The inner planet's $e$ and $i$ vary from 0 to
0.75 and $55^\circ$, respectively.  After $\sim 6$~Gyr, the system
slowly moves into a different state with $i_1$ varying between
$20^\circ$ and $50^\circ$, and $\theta_1$ aperiodically circulating.
System F has resonant arguments that switch between libration and
circulation as in Fig.~\ref{fig:105_100kyr} and eccentricities that
remain below 0.3 and inclinations below $20^\circ$. System G is
similar to System A.

These examples are only illustrative and do not represent the full
range of possibilities. The 7 cases listed in Table~\ref{tab:set1} met our
stability criteria (see $\S$~\ref{sec:methods}), and we suspect many
more cases also would, but computational constraints prevented a more
thorough analysis.

\subsubsection{Two Giant Planets Orbiting a 0.75~$\msun$ Star (Set \#2)}

Next we consider Set \#2: Two Saturn- to Jupiter-mass planets orbiting a 0.75~$\msun$
star in 2 and 4 year orbital periods. Such planets induce a larger
astrometric signal in the host star, and are large enough that radial
velocity observations could break the 180$^\circ$ ambiguity in $\Omega$
implicit in astrometric measurements of exoplanets. In Table~\ref{tab:set2} we present 7 systems which survived for $\sim10$~Gyr and conserved energy adequately.

\begin{deluxetable*}{ccccccccc}
\tablecolumns{9}
\tablecaption{Initial Conditions for Selected Set \#2 Systems}
\tabletypesize{\normalsize}
\tablehead{\colhead{System} & \colhead{Body} & \colhead{$m$ ($M_{Jup}$)} & 
	\colhead{$a$ (AU)} & \colhead{$e$} & \colhead{$i$ ($^\circ$)} & 
	\colhead{$\Omega$ ($^\circ$)} & \colhead{$\omega$ ($^\circ$)} & 
	\colhead{$\mu$ ($^\circ$)}}
\startdata
H & 1 & 0.314 & 1.4422 & 0.0276 & 2.504 & 247.16 & 94.57 & 7.74\\     
  & 2 & 0.573 & 2.2893 & 0.2021 & 1.118 & 67.12 & 126.15 & 313.87\\
I$^e$ & 1 & 0.314 & 1.4422 & 0.1155 & 3.302 & 39.32 & 273.88 & 195.14\\   
  & 2 & 0.303 & 2.2893 & 0.4692 & 3.076 & 219.29 & 247.62 & 48.35\\
J & 1 & 0.314 & 1.4422 & 0.2374 & 7.089 & 24.56 & 256 & 77.1\\        
  & 2 & 0.234 & 2.2893 & 0.3988 & 8.019 & 204.54 & 185.86 & 127.68\\
K & 1 & 0.314 & 1.4422 & 0.0364 & 11.267 & 126.11 & 284.5 & 41.42\\   
  & 2 & 0.328 & 2.2893 & 0.223 & 8.761 & 306.18 & 273.49 & 160.47\\
L & 1 & 0.314 & 1.4422 & 0.0263 & 19.459 & 176.25 & 202.97 & 229.91\\ 
  & 2 & 0.736 & 2.2893 & 0.3696 & 6.989 & 356.23 & 153.08 & 261.74\\
M & 1 & 0.314 & 1.4422 & 0.4596 & 3.282 & 106.68 & 44.59 & 350\\      
  & 2 & 0.399 & 2.2893 & 0.3402 & 1.936 & 286.57 & 131.7 & 73.93\\
N & 1 & 0.314 & 1.4422 & 0.0561 & 21.623 & 247.55 & 325.64 & 274.59\\ 
  & 2 & 0.902 & 2.2893 & 0.4466 & 6.53 & 67.64 & 60.26 & 326.55\\
\tablenotetext{e}{Stable for only 9.761~Gyr.}
\label{tab:set2}
\end{deluxetable*}

\begin{figure*}
\begin{tabular}{cc}
\includegraphics[width=0.49\textwidth]{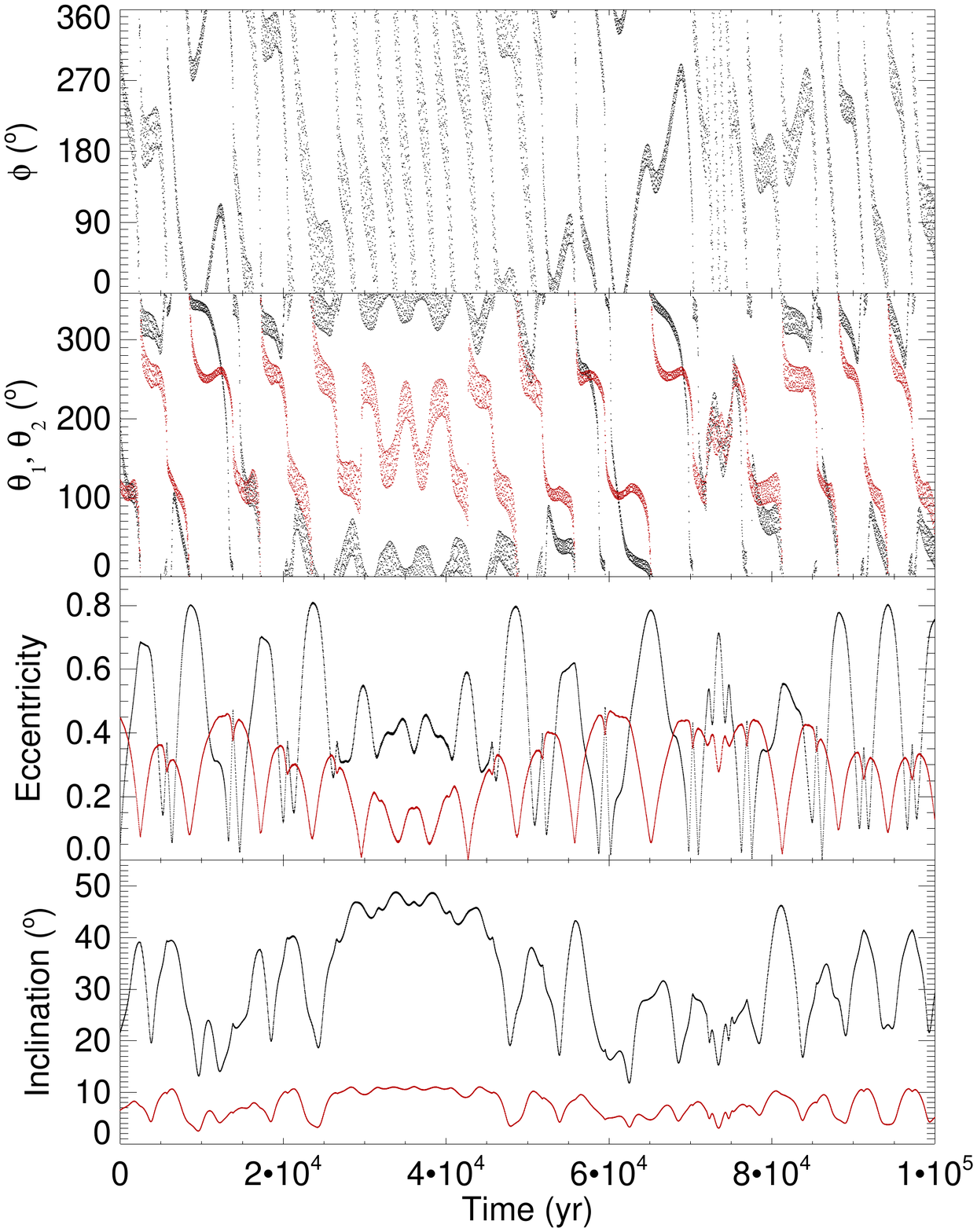} &
\includegraphics[width=0.49\textwidth]{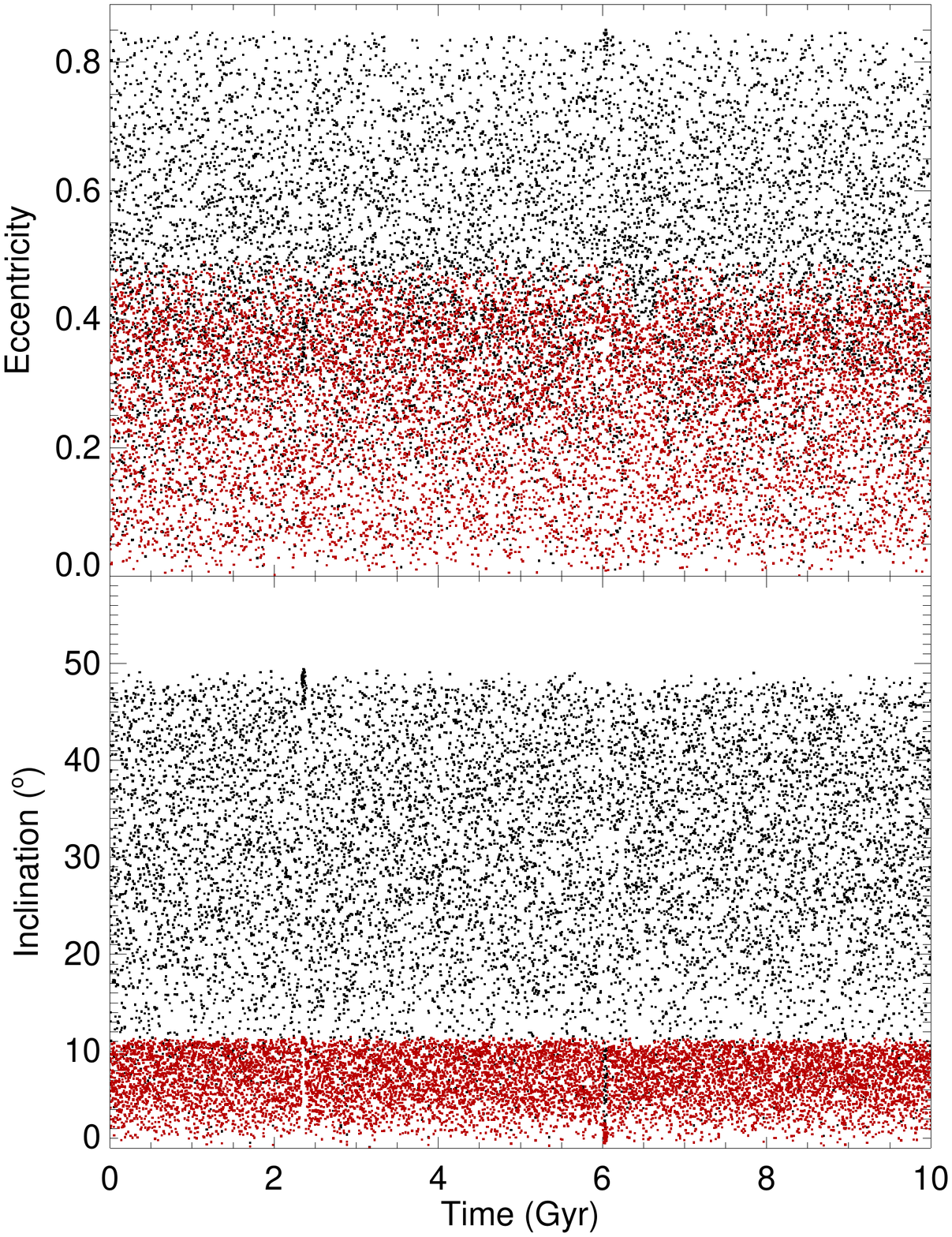}
\end{tabular}
\caption{Orbital evolution of System N. The left panels are in the same format as Fig.~\ref{fig:105_100kyr}; the right is in the same format as Fig.~\ref{fig:105_10Gyr}.}
\label{fig:TS_191}
\end{figure*}

In Fig.~\ref{fig:TS_191} we show the evolution of System N on two
timescales. The left panels show the variation in the resonant
arguments and $e$ and $i$ for $10^5$ years in the same format as
Fig.~\ref{fig:105_100kyr}. As before the resonant arguments switch
between libration and circulation leading to chaotic evolution of $e$
and $i$. In the right panels we show the evolution of $e$ and $i$ for
10~Gyr. $e_1$ aperiodically reaches values of $\sim~0.85$, and $i_1$ reaches
$50^\circ$. Note as well at 2.4 and 6.0 Gyr the systems enters 
qualitatively different states for about 100~Myr. 

\begin{figure*} 
\includegraphics[width=0.99\textwidth]{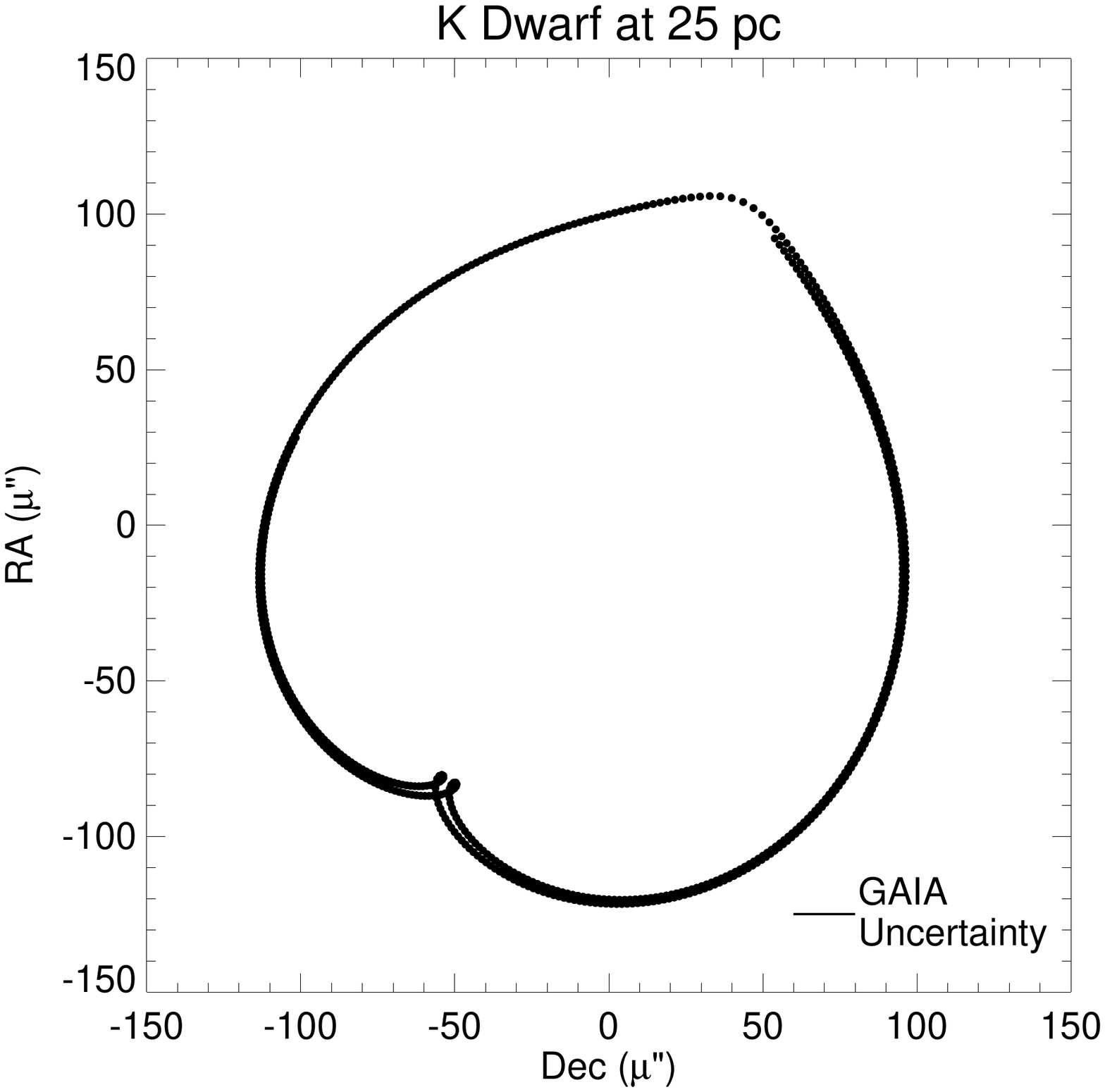}
\caption{Orbit of the host star in Fig.~\ref{fig:TS_191} (System N) about the system barycenter for 7 years if the system lies at 25~pc and the fundamental plane is parallel to the sky plane. The line labeled ``GAIA Uncertainty'' is 25 $\mu$as long, and represents {\it GAIA}'s typical per-measurement uncertainty for bright stars.} 
\label{fig:astrom}
\end{figure*}

In Fig.~\ref{fig:astrom} the orbit of the host star about the system's
barycenter is shown over 7 years if viewed face-on, \ie the invariable
plane is parallel to the sky plane. In this example we assume the
system is located at 25~pc, and the combined system induces a $\sim
100$ $\mu$as astrometric signal, which is 4--5 times larger than the
expected {\it GAIA} uncertainties for stars with G-band magnitudes
$\lsim 13$, represented by the line labeled ``GAIA Uncertainty''
\citep{Sozzetti14}. This system would be relatively easy to
characterize with radial velocity data as well, and hence could be
fully characterized in the next 10 years. Of course, the details of
actually modeling resonant systems are non-trivial
\citep[\eg][]{Marcy01}, but the discovery of such a chaotic system would mark an
important milestone in exoplanet science.

\begin{figure*}
\begin{tabular}{cc}
\includegraphics[width=0.49\textwidth]{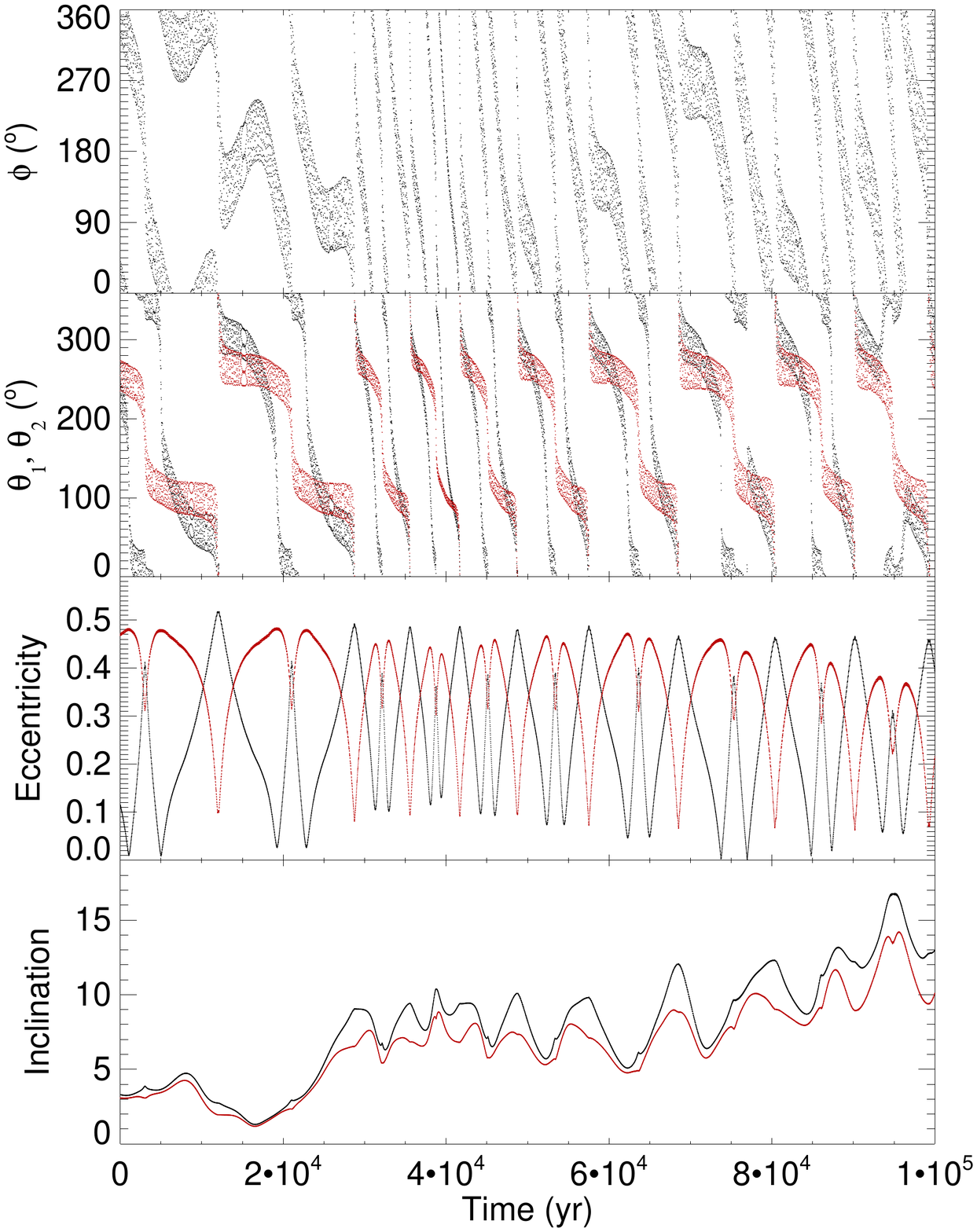} &
\includegraphics[width=0.49\textwidth]{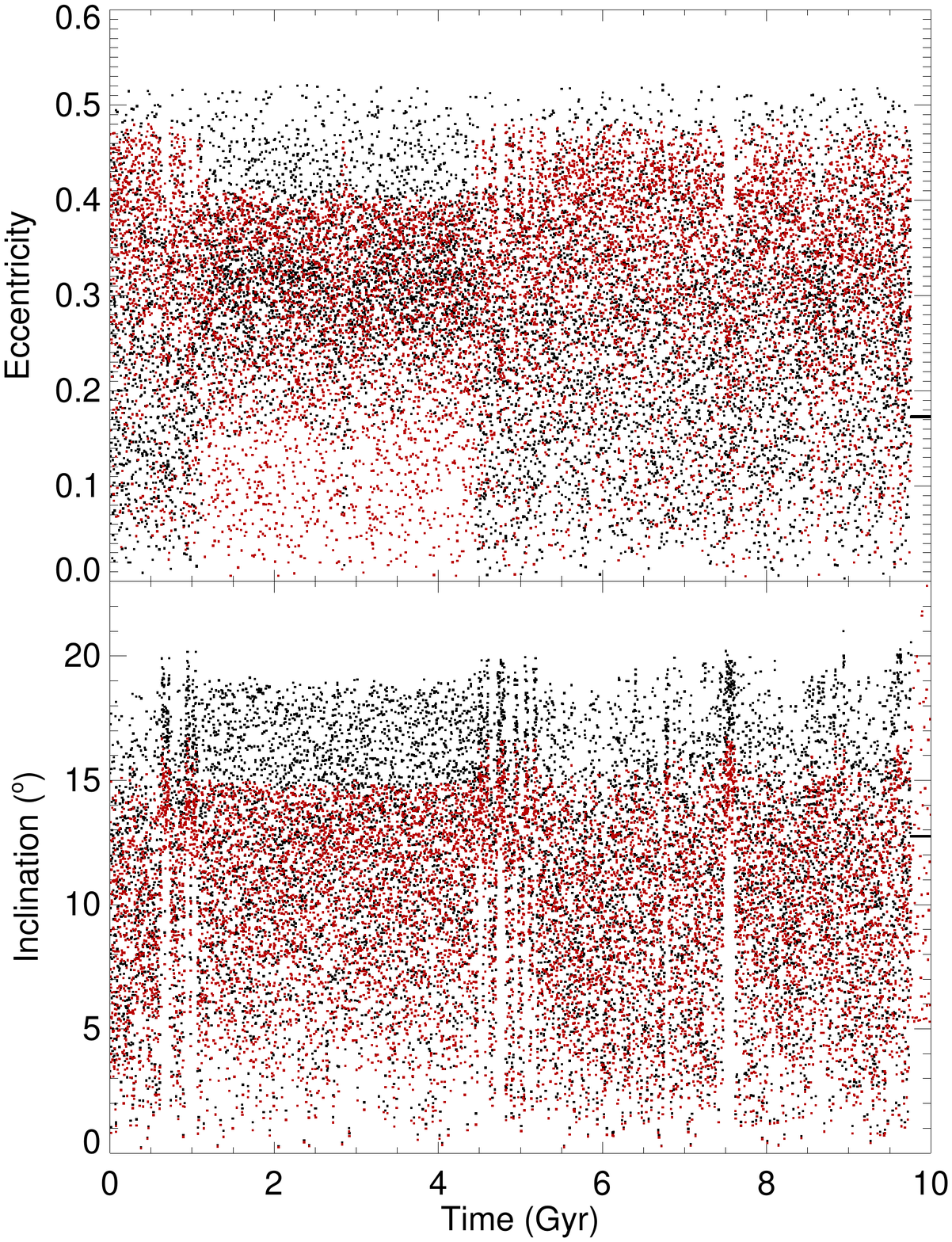}
\end{tabular}
\caption{Orbital evolution of System I in the same format as Fig.~\ref{fig:TS_191}. The system destabilizes after 9.761~Gyr.}
\label{fig:TS_143}
\end{figure*}

In Fig.~\ref{fig:TS_143} we show the evolution of System I in the same
format as Fig.~\ref{fig:TS_191}. This system appears qualitatively
similar to System N, but with lower amplitudes in $e$ and
$i$. However, this system destabilizes at 9.761~Gyr, as seen on the
right side of the right panels. We found several other systems in
which an ejection occurred after 1~Gyr. The chaotic resonant behavior
shown in this study is not necessarily stable on arbitrarily long
timescales. Like our own Solar System, these hypothetical systems are
just relatively long-lived
\cite[see, \eg][]{Lecar01}.

\subsubsection{Dwarf Planets in the 2:1 MMR (Set \#3)}

\begin{figure*}
\includegraphics[width=0.99\textwidth]{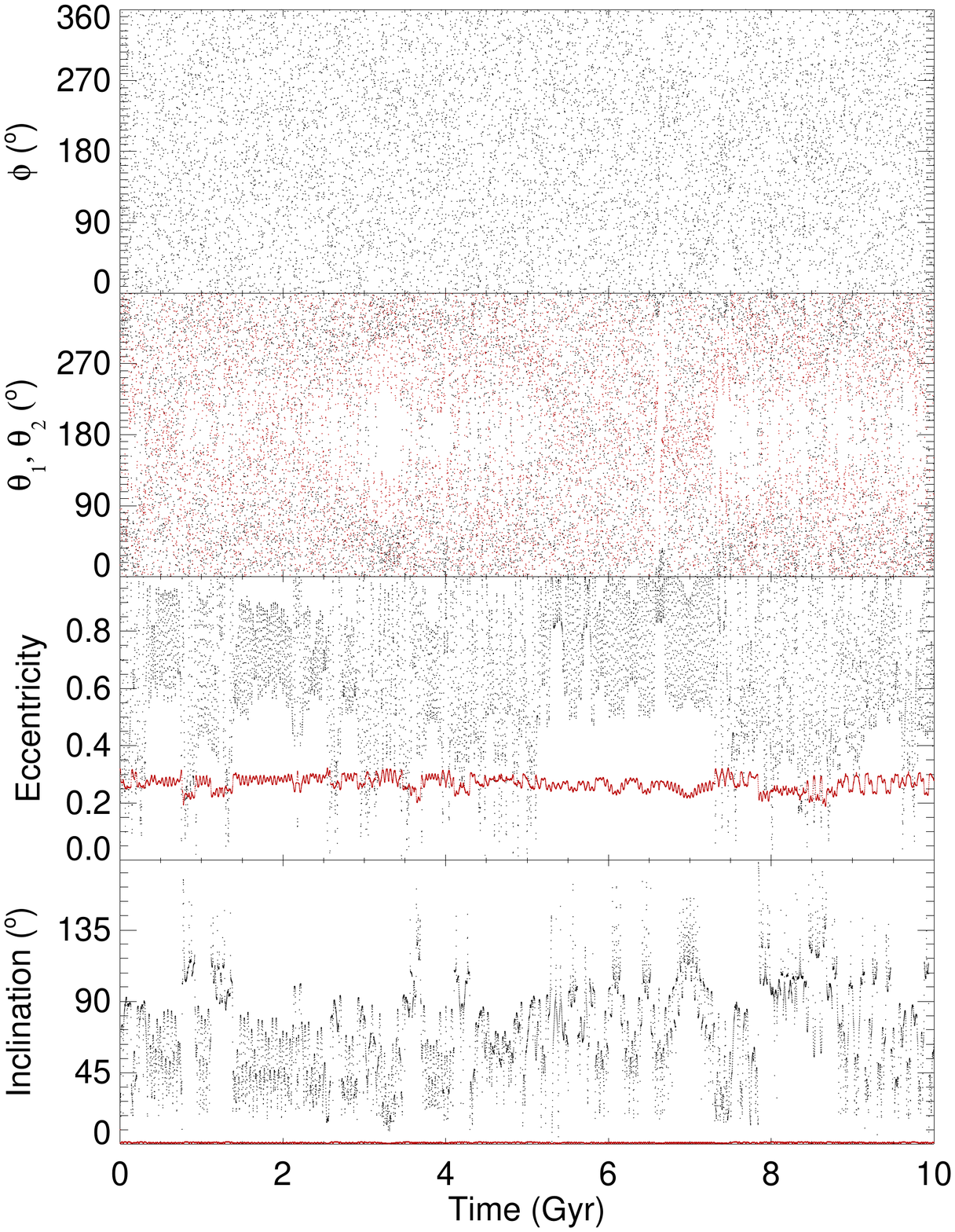}
\caption{Evolution of System P in the same format as Fig.~\ref{fig:105_100kyr}.}
\label{fig:175}
\end{figure*}

In this section we consider Set \#3, dwarf planets in the 2:1
MMR. While these planets are not detectable in the near future, they
display remarkable dynamics and illustrate the extreme chaos that the
2:1 MMR with inclinations can produce. In Table~\ref{tab:set3} we present 4 cases
that are stable for 10~Gyr, and in Fig.~\ref{fig:175} we show the
evolution of System P over 10 Gyr. Note that the $e$-resonance
arguments sometimes librate for long periods of time, such as near 2.5
Gyr. The $i$ argument does not appear to librate in this visualization,
but higher resolution plots over shorter periods show similar behavior
as above. Systems O and Q are similar to P, but the inclinations and
eccentricities remain lower.

\begin{deluxetable*}{ccccccccc}
\tablecaption{Initial Conditions for Selected Set \#3 Systems}
\tabletypesize{\normalsize}
\tablecolumns{9}
\tablehead{\colhead{System} & \colhead{Body} & \colhead{$m$ (M$_{Moon}$)} & 
	\colhead{$a$ (AU)} & \colhead{$e$} & \colhead{$i$ ($^\circ$)} & 
	\colhead{$\Omega$ ($^\circ$)} & \colhead{$\omega$ ($^\circ$)} & 
	\colhead{$\mu$ ($^\circ$)}}
\startdata
O & 1 & 0.0775 & 1 & 0.3032 & 19.835 & 283.31 & 290.53 & 231.72\\     
  & 2 & 1.171 & 1.5874 & 0.2768 & 1.013 & 103.31 & 50.74 & 54.65\\
P & 1 & 0.0775 & 1 & 0.04305 & 24.34 & 57.06 & 90.49 & 57.09\\        
  & 2 & 2.817 & 1.5874 & 0.3192 & 0.544 & 237.06 & 231.06 & 90.68\\
Q & 1 & 0.0775 & 1 & 0.1036 & 5.298 & 169.03 & 21.27 & 348.68\\       
  & 2 & 0.568 & 1.5784 & 0.1544 & 0.577 & 349.03 & 171.19 & 158.69\\
R & 1 & 0.0775 & 1 & 0.4475 & 14.98 & 289.05 & 94.15 & 272.54\\       
  & 2 & 2.513 & 1.5784 & 0.2093 & 0.332 & 109.05 & 352.99 & 154.19\\
\label{tab:set3}
\end{deluxetable*}

These systems show a diversity of behavior from small-scale chaos
(System Q), to dramatic chaos in which $e_1$ and $i_1$ sample all
available phase space (System P). The resonant arguments, particularly
the eccentricity arguments, switch between circulation and
libration. $e_1$ in Systems O (not shown),
P and R (not shown) aperiodically reach values in excess of 0.99, and
hence they should tidally circularize prior to 10~Gyr.

The resonant arguments in these systems behave differently than for
the larger planets. On short timescales (not shown) the resonant arguments switch modes as in previous cases, but these systems can remain in one mode for long timescales, particularly the $e$-resonance. In
Fig.~\ref{fig:175} note that there are intervals when the two
$e$-arguments librate for more than 100~Myr.

\subsection{The 3:1 Resonance}

In this section we consider an Earth-like planet at 1 AU from a $1~\msun$ star with an exterior companion with an orbital period of 3 years, \ie in the 3:1 MMR.  In this case the $e$-resonance arguments are 
\begin{equation}
\theta_{1,2} = 3\lambda_2 - \lambda_1 - 2\varpi_{1,2},
\label{eq:e_res_3:1}
\end{equation}
and the $i$-resonance argument is
\begin{equation}
\phi = 3\lambda_2 - \lambda_1 - \Omega_1 - \Omega_2.
\label{eq:i_res_3:1}
\end{equation}

\begin{figure*}[h] 
\begin{tabular}{cc}
\includegraphics[width=0.49\textwidth]{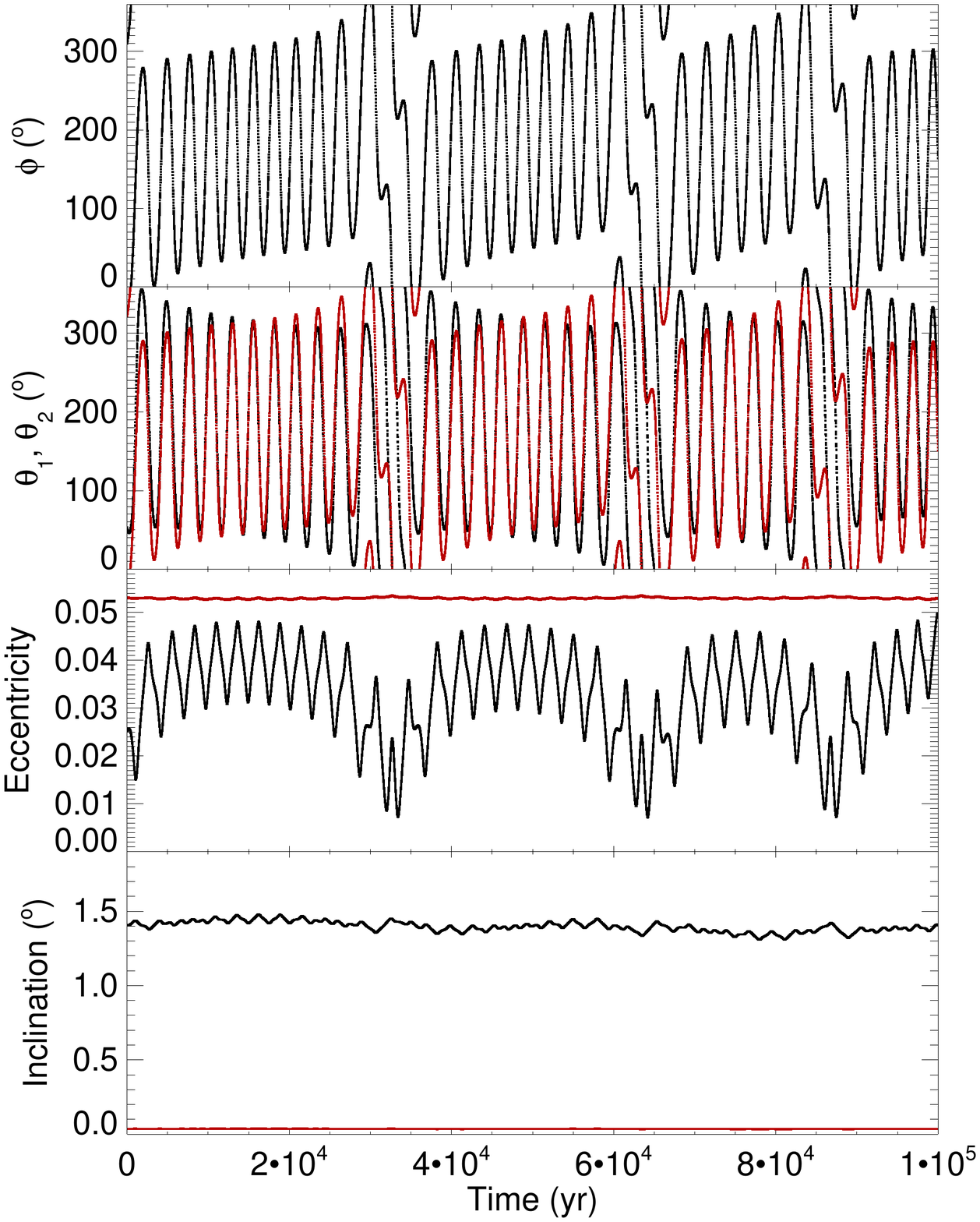} &
\includegraphics[width=0.49\textwidth]{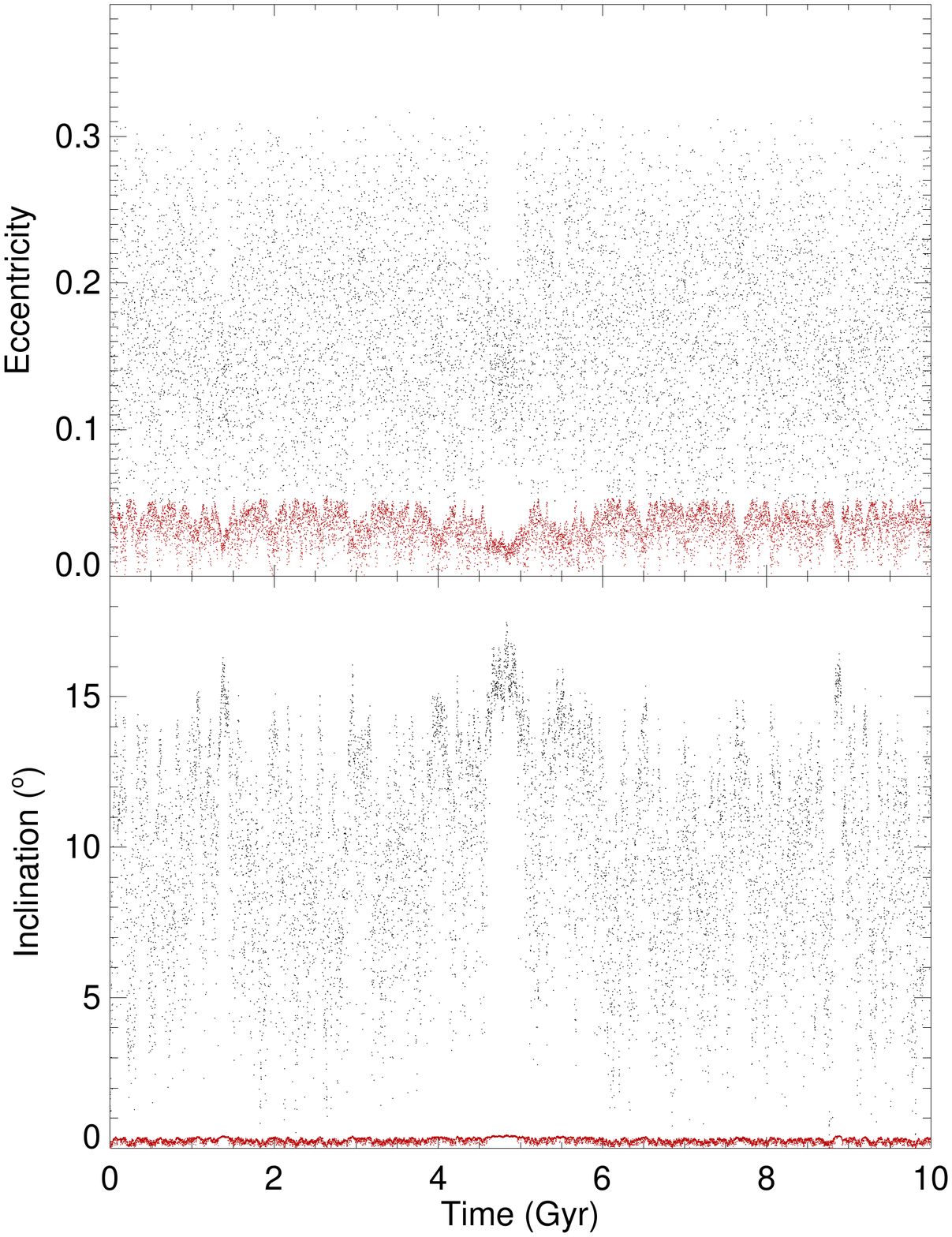}
\end{tabular}
\caption{Orbital evolution of System T in the same format as Fig.~\ref{fig:TS_191}.}
\label{fig:3:1_104}
\end{figure*}

\begin{figure*}
\begin{tabular}{cc}
\includegraphics[width=0.49\textwidth]{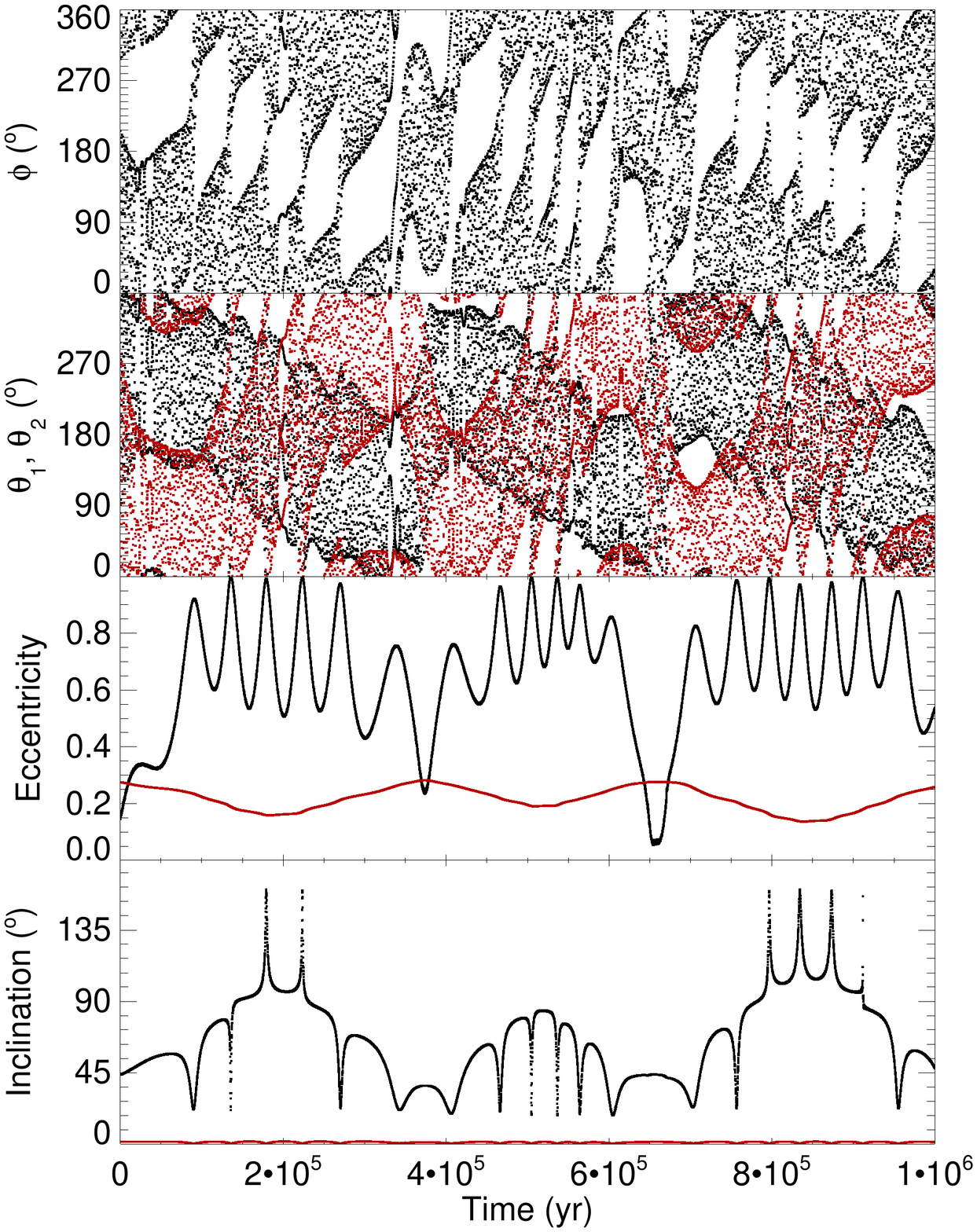} &
\includegraphics[width=0.49\textwidth]{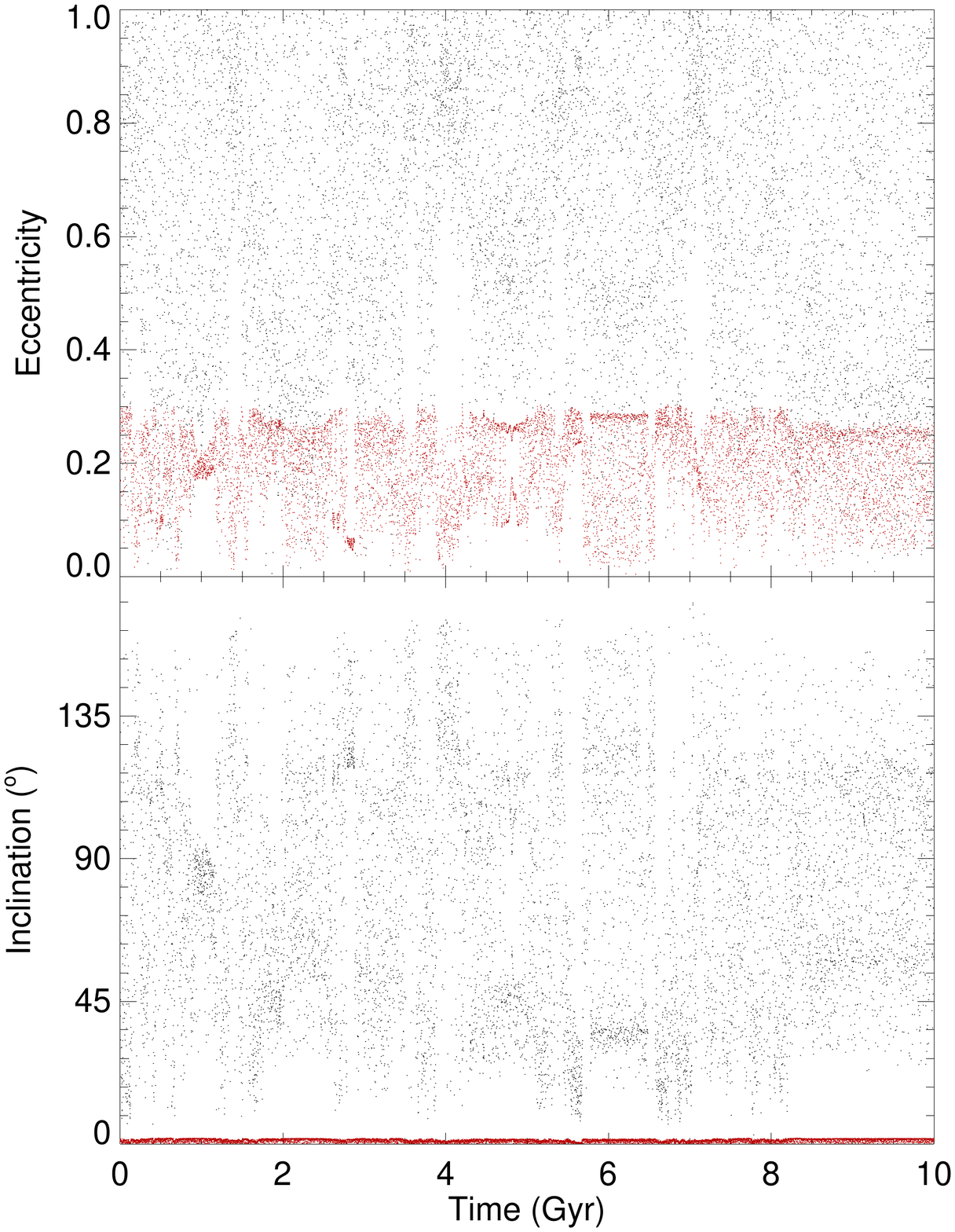}
\end{tabular}
\caption{Orbital evolution of System U in the same format as Fig.~\ref{fig:TS_191}.}
\label{fig:3:1_107}
\end{figure*}

Table~\ref{tab:set4} lists three configurations that are stable for 10~Gyr and show
chaotic evolution. Systems T and U are shown in
Figs.~\ref{fig:3:1_104} and \ref{fig:3:1_107}, respectively. The
former has (relatively) modest inclination and eccentricity
variations, while the latter samples all the phase space available,
with $0~\lsim~i~\lsim~\pi$ and $0~\lsim~e~\lsim~1$. In both systems the resonance arguments switch between libration and circulation, as seen in the previous 2:1 MMR cases. System S (not
shown) is similar to System U, but the inclinations only reach
$\sim~65^\circ$ and $e$ only 0.9.

\begin{deluxetable*}{ccccccccc}
\tablecaption{Initial Conditions for Selected Set \#4 Systems}
\tablecolumns{9}
\tablehead{\colhead{System} & \colhead{Body} & \colhead{$m$ (M$_\oplus$)} & 
	\colhead{$a$ (AU)} & \colhead{$e$} & \colhead{$i$ ($^\circ$)} & 
	\colhead{$\Omega$ ($^\circ$)} & \colhead{$\omega$ ($^\circ$)} & 
	\colhead{$\mu$ ($^\circ$)}}
\startdata
S & 1 & 1 & 1 & 0.1798 & 2.06 & 178.04 & 261.79 & 66.85\\            
  & 2 & 6.5 & 2.0801 & 0.3808 & 0.234 & 358.4 & 113.3 & 286.96\\
T & 1 & 1 & 1 & 0.02514 & 1.04 & 246.96 & 151.63 & 261.33\\          
  & 2 & 27.05 & 2.0801 & 0.05311 & 0.0362 & 66.96 & 17.04 & 359.38\\
U & 1 & 1 & 1 & 0.149 & 43.64 & 260.51 & 18.15 & 220.4\\             
  & 2 & 22.2 & 2.0801 & 0.2755 & 1.27 & 80.51 & 51.82 & 6.72
\label{tab:set4}
\end{deluxetable*}

\subsection{The 3:2 Resonance}

Next we consider the 3:2 MMR with an interior
$1~\mearth$ planet at 1~AU from a solar-mass star and a larger
external companion at 1.3104~AU. The $e$-resonance arguments are
\begin{equation}
\theta_{1,2} = 3\lambda_2 - 2\lambda_1 - \varpi_{1,2},
\label{eq:e_res_3:2}
\end{equation}
and the $i$-resonance argument is
\begin{equation}
\phi = 6\lambda_2 - 4\lambda_1 - \Omega_1 - \Omega_2.
\label{eq:i_res_prime_3:2}
\end{equation}

\begin{figure*}[h] 
\begin{tabular}{cc}
\includegraphics[width=0.49\textwidth]{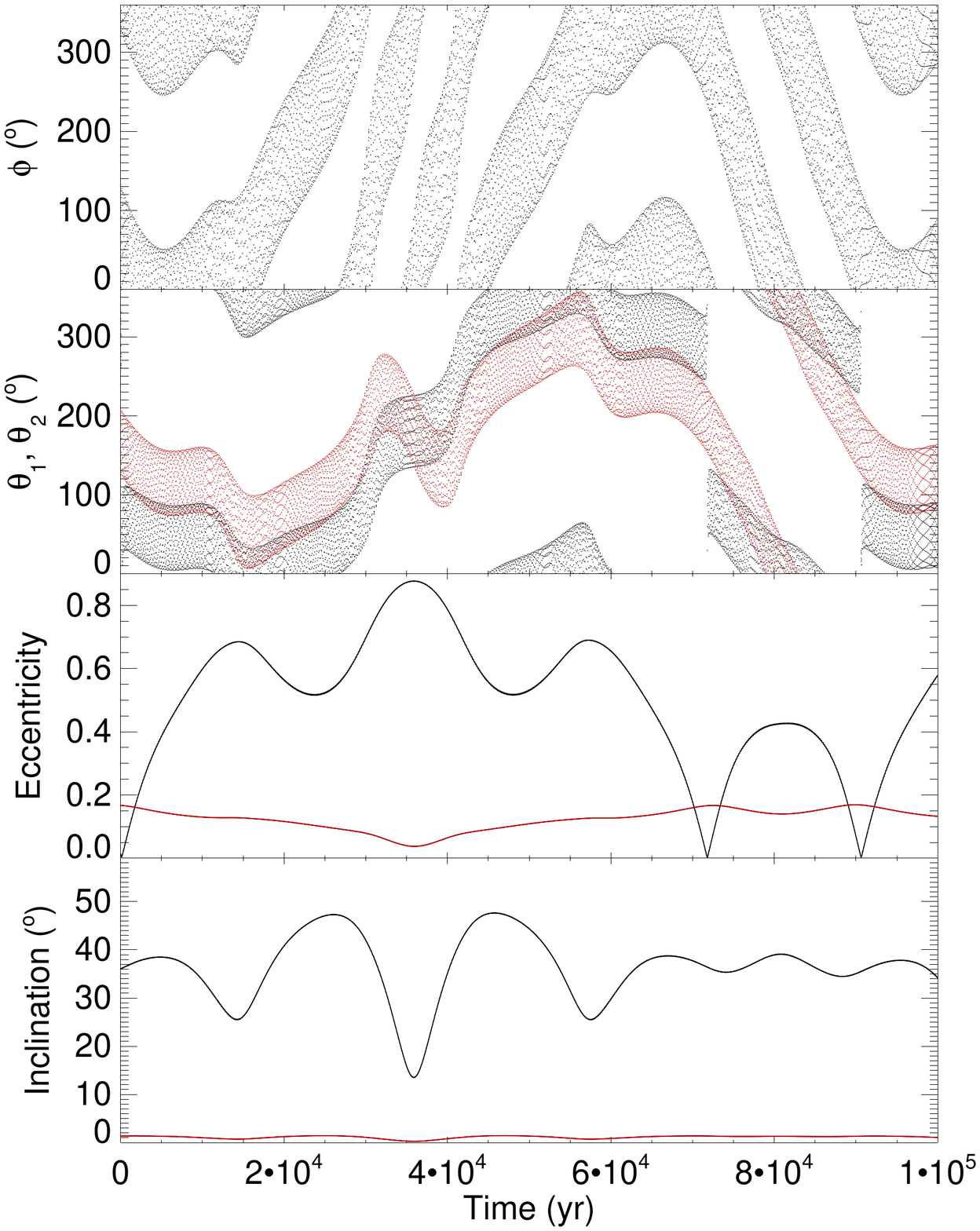} &
\includegraphics[width=0.49\textwidth]{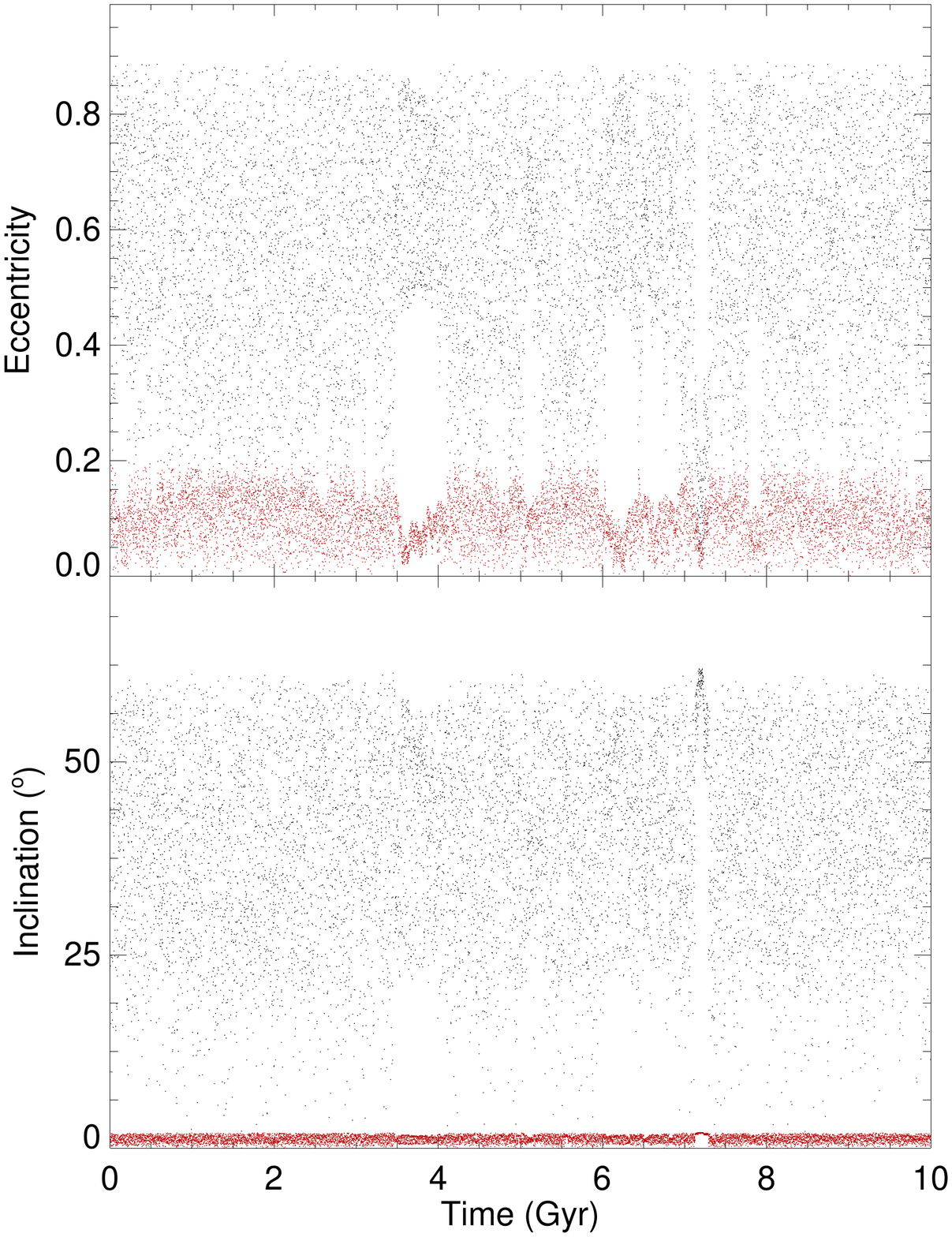}
\end{tabular}
\caption{Orbital evolution of System W in the same format as Fig.~\ref{fig:TS_191}.}
\label{fig:3:2_121}
\end{figure*}

Table~\ref{tab:set5} lists two systems that are stable for 10~Gyr and showed chaotic
evolution. In Fig.~\ref{fig:3:2_121} we plot the evolution of the
resonant arguments, $e$ and $i$ on two timescales for System W. The
evolution is qualitatively similar to that in the other
resonances. System V is quite similar, with $e$ reaching 0.9 and $i$
reaching $60^\circ$ aperiodically over 10~Gyr.

\begin{deluxetable*}{ccccccccc}
\tablecaption{Initial Conditions for Selected Set \#5 Systems}
\tablecolumns{9}
\tablehead{\colhead{System} & \colhead{Body} & \colhead{$m$ (M$_\oplus$)} & 
	\colhead{$a$ (AU)} & \colhead{$e$} & \colhead{$i$ ($^\circ$)} & 
	\colhead{$\Omega$ ($^\circ$)} & \colhead{$\omega$ ($^\circ$)} & 
	\colhead{$\mu$ ($^\circ$)}}
\startdata
V & 1 & 1 & 1 & 0.1404 & 27.34 & 232.78 & 39.79 & 5.92\\             
  & 2 & 5.96 & 1.3104 & 0.3498 & 4.08 & 52.78 & 314.14 & 108.38\\
W & 1 & 1 & 1 & 0.0144 & 36.05 & 316.73 & 13.72 & 293.02\\           
  & 2 & 21.83 & 1.3104 & 0.1671 & 1.37 & 136.73 & 307.8 & 308.84\\
\label{tab:set5}
\end{deluxetable*}

\section{Formation by Scattering}
\label{sec:form}

The large amplitude chaotic orbital evolution shown above is
remarkable, but can such a system form? As described in
$\S$~\ref{sec:intro}, several studies have examined the formation of
inclination resonances in the context of convergent migration during
the protoplanetary disk phase. Those studies found that $i$-resonances
can form under the proper circumstances, but they did not consider the
post-formation evolution. In this section, we show that gravitational
scattering among planets can produce systems in the 2:1 MMR with mutual
inclinations that evolve chaotically for 10~Gyr.

Our sample comes from the data set used in \cite{Raymond08} that found
MMRs resulting from scattering events. The reader is referred to that
paper for details, but briefly, systems consisting of initially 3
planets were allowed to interact gravitationally and in many cases
1--2 planets were ejected from the system. \cite{Raymond08} examined
the $e$-resonance arguments of systems in which 2 planets remained,
and reported that 5\% of systems had at least one $e$-resonance
argument librating. Systems like those shown in $\S$~\ref{sec:hypo}
were deemed unstable and thrown out.

\begin{figure*}
\includegraphics[width=0.99\textwidth]{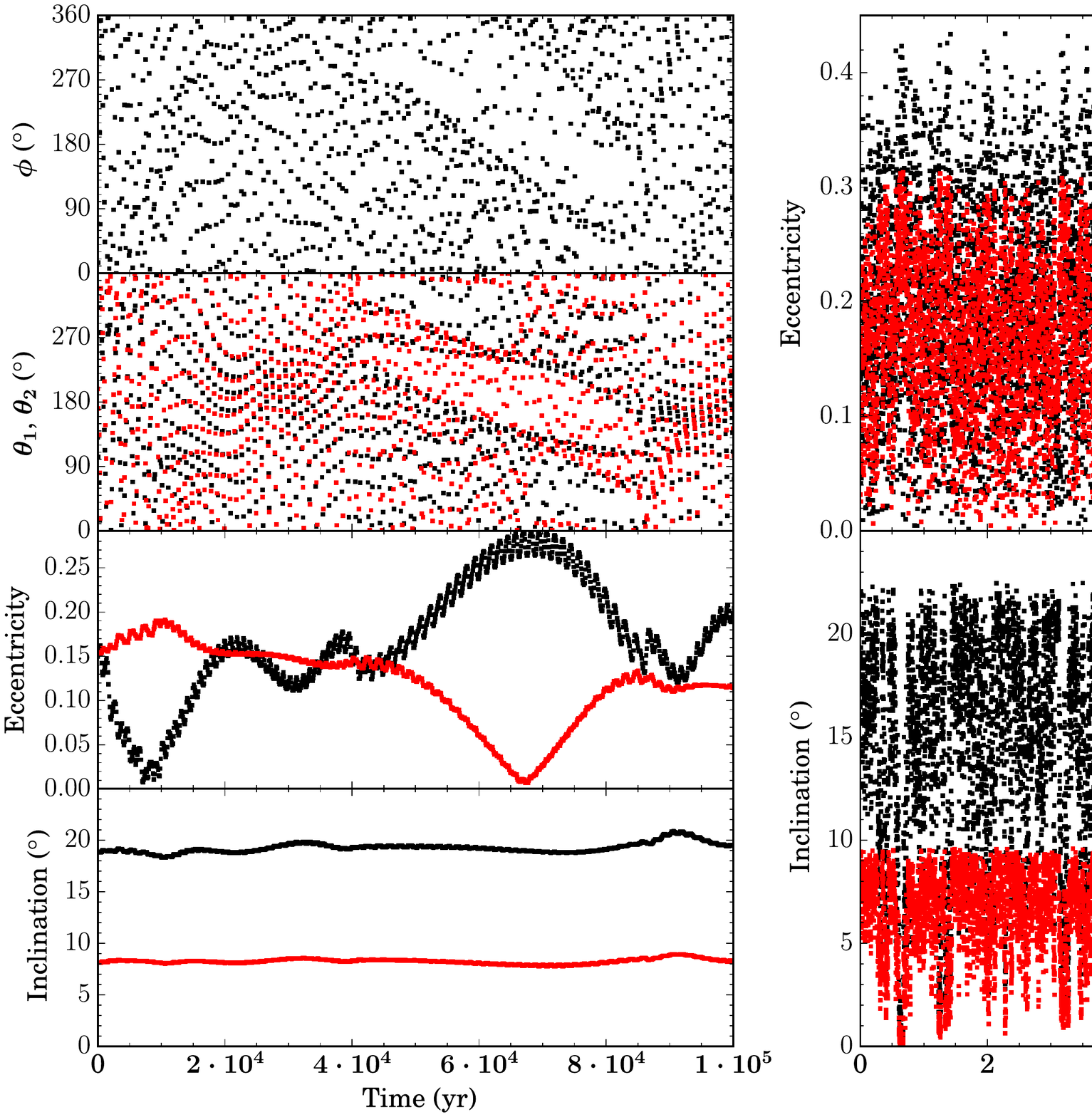}
\caption{Orbital evolution of System X in the same format as Fig.~\ref{fig:TS_191}.}
\label{fig:scat383}
\end{figure*}

In light of the results of $\S$~\ref{sec:hypo} we have re-examined
systems near an MMR to search for chaotic, but long-lived
evolution. Specifically we focus on the ``Mixed2'' distribution of
\cite{Raymond08} in which the planets' masses follow a power-law
distribution with exponent $-1.1$. This distribution has recently been
shown to reproduce many observed dynamical properties of exoplanets
\citep{Timpe13}. This suite of simulations consisted of 1000 systems,
and we examined 49 that had period ratios with 10\% of 2, of which 24 were
identified as being in an $e$-resonance by \cite{Raymond08}. Of these,
27 had mutual inclinations less than $5^\circ$ and the largest was
27$^\circ$. We find three that appear qualitatively similar to those
in the previous section. They are listed in Table~\ref{tab:form}, and System X is
shown in Fig.~\ref{fig:scat383}. System Y (not shown) evolved such
that the eccentricities and inclinations remained below 0.12 and
$7^\circ$, respectively. System Z evolved such that they remained
below 0.15 and $3^\circ$. We also searched for systems evolving
chaotically in the 3:1 or 3:2 MMR, but did not find any.

The amplitudes of the variations of the orbital elements is lower than
some cases shown in $\S$~\ref{sec:hypo}, but similar to others, \eg
System B. Given the small number of systems that produced chaotic
resonant behavior, it remains to be seen if evolution that reaches $e~\sim~1$
and $i~\sim~\pi$ can be naturally produced. Nonetheless we conclude
that planet-planet scattering can produce systems that evolve
chaotically for 10~Gyr in an MMR.

\begin{deluxetable*}{ccccccccc}
\tablecaption{Initial Conditions for Planets in a Chaotic 2:1 MMR Formed by Scattering}
\tablecolumns{9}
\tablehead{\colhead{System} & \colhead{Body} & \colhead{$m$ (M$_{Jup}$)} & 
	\colhead{$a$ (AU)} & \colhead{$e$} & \colhead{$i$ ($^\circ$)} & 
	\colhead{$\Omega$ ($^\circ$)} & \colhead{$\omega$ ($^\circ$)} & 
	\colhead{$\mu$ ($^\circ$)}}
\startdata
X & 1 & 0.118 & 4.83696 & 0.1498 & 18.845 & 291.14 & 83.60 & 330.83 \\   
  & 2 & 0.213 & 7.71971 & 0.1548 & 8.15 & 111.12 & 195.95 & 118.57 \\
Y & 1 & 0.65 & 6.23933 & 0.026033 & 0.802 & 168.16 & 30.05 & 284.95 \\   
  & 2 & 0.0737 & 9.97199 & 0.04723 & 5.60 & 348.18 & 298.78 & 49.35 \\
Z & 1 & 0.168 & 5.45138 & 0.13299 & 1.453 & 133.88 & 186.20 & 113.99 \\  
  & 2 & 0.934 & 8.78422 & 0.01305 & 0.204 & 313.83 & 137.53 & 146.93 \\
\label{tab:form}
\end{deluxetable*}

\section{Known Systems}
\label{sec:known}

The previous two sections established the typical characteristics of
planets in inclined MMRs and a viable formation mechanism. The next
question that naturally arises is if any known systems might be
evolving in a long-lived and chaotic configuration. In this section we
examine four candidates in 3 different MMRs and with observed
properties listed in Table~\ref{tab:known}. These planets were all discovered by the
radial velocity technique and hence their orbital planes were
undetected. HD~128311~c has been detected astrometrically
\citep{McArthur14}, but its companion planet has not, so the mutual
inclination remains unknown.

\subsection{2:1 Systems: HD~128311 and HD~73426}

We performed 100 integrations of HD~128311 and HD~73526 for 10~Myr
each. We found no stable configurations of the former if the orbits
were allowed to be non-planar. This should not be taken to mean the
system must be coplanar as we may not have considered enough
cases. \cite{ReinPapaloizou09} found the system could form in a
coplanar configuration through convergent migration, and
\cite{McArthur14} found the best-fit coplanar solution to the system
is dynamically stable and not in resonance. At this point, we conclude that this system is
likely to be in a coplanar configuration, which precludes the chaotic
evolution we report here.

\begin{figure*}
\begin{tabular}{cc}
\includegraphics[width=0.49\textwidth]{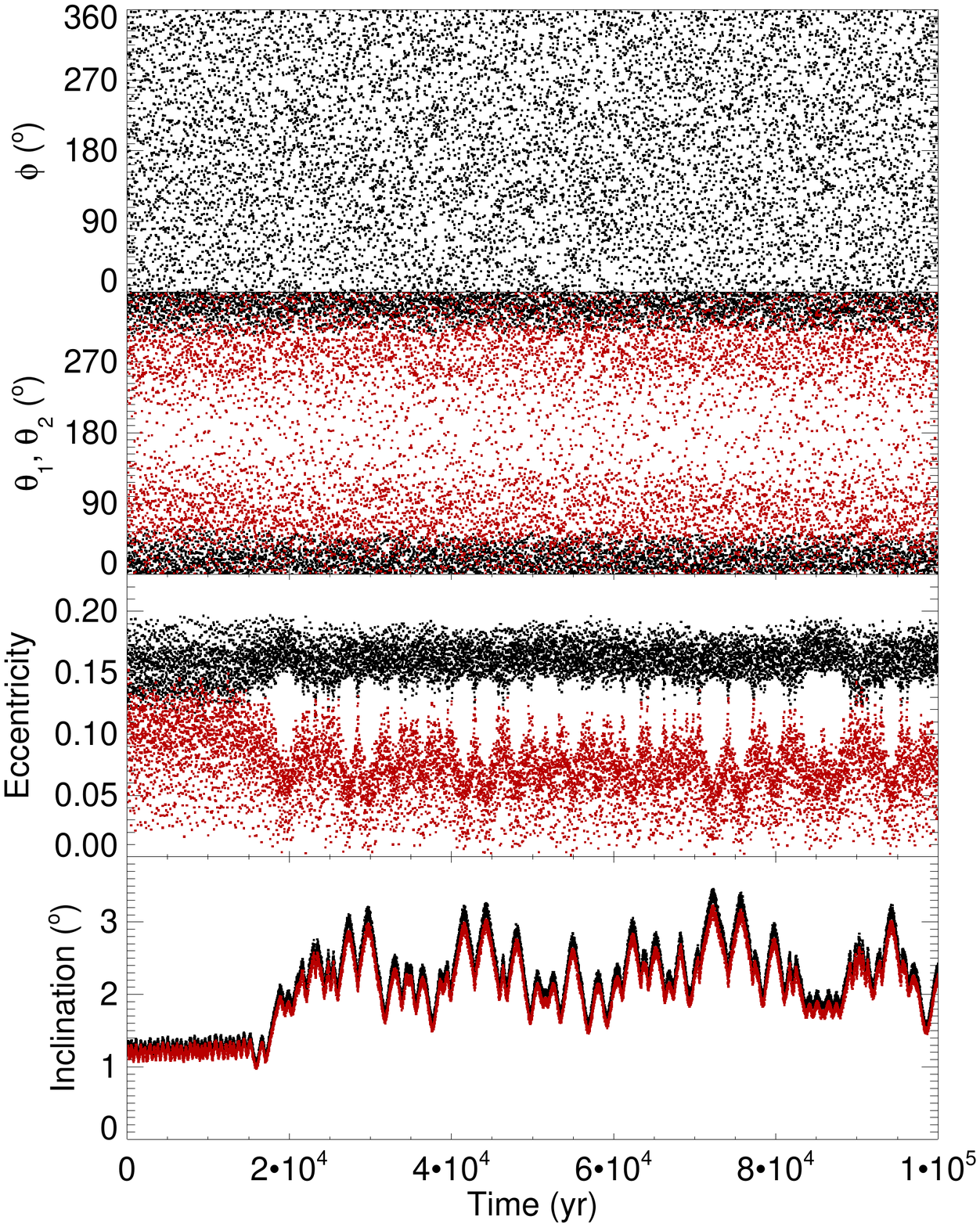} &
\includegraphics[width=0.49\textwidth]{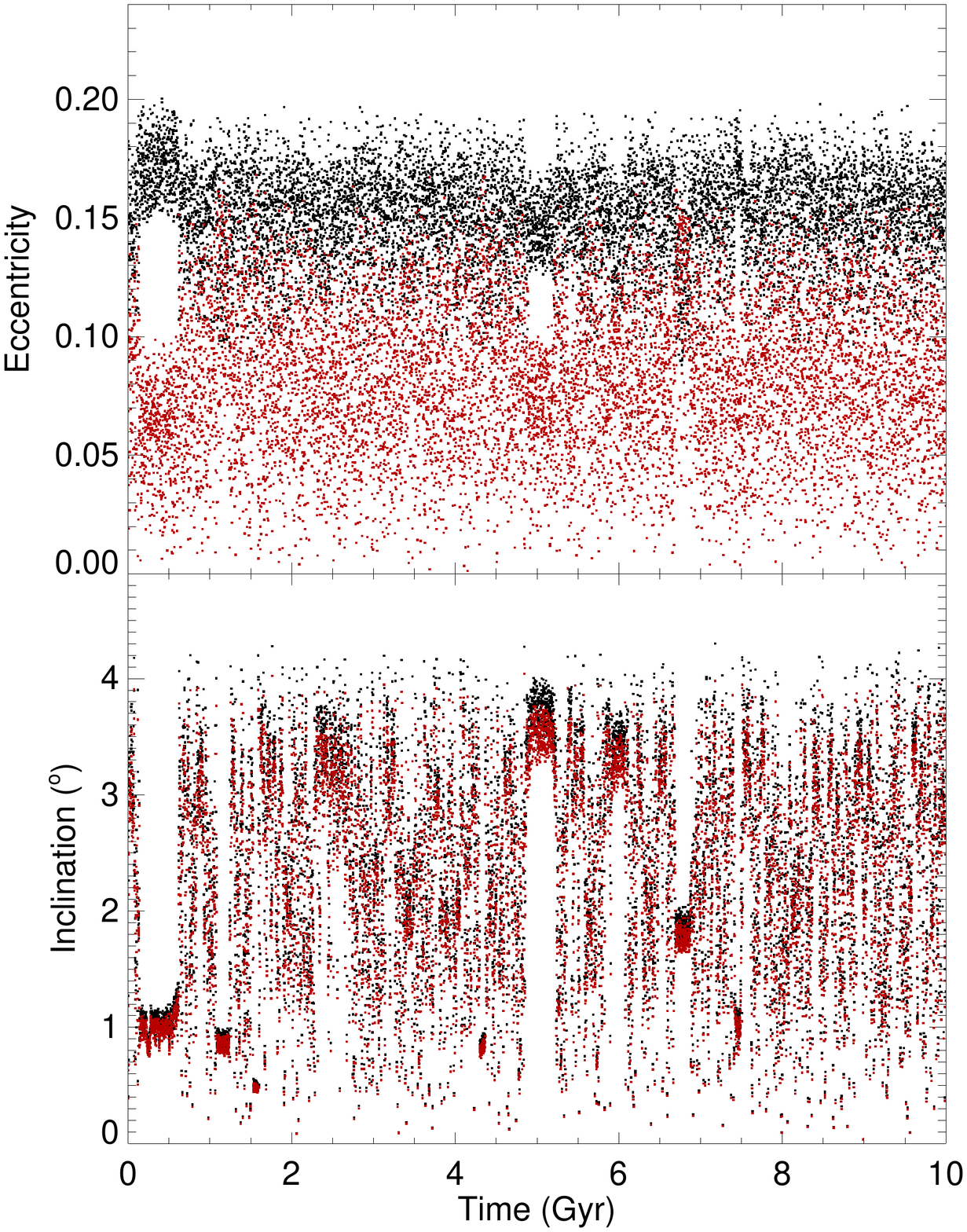}
\end{tabular}
\caption{Orbital evolution of System AB (HD~73526) in the same format as Fig.~\ref{fig:TS_191}.}
\label{fig:hd73526}
\end{figure*}

HD~73526, on the other hand, could be evolving chaotically. In Table~\ref{tab:hd73526}
we list three versions that are stable for 10~Gyr and show chaotic
evolution. Note that we always use a stellar mass of $1.08~\msun$. In
Fig.~\ref{fig:hd73526} we show the evolution of System AB. Unlike the
previous systems, the resonant arguments do not appear to switch
between libration and circulation, and the $i$-arguments do not show
any libration at all. The $\theta_1$ argument librates about 0 for the
full 10 Gyr year duration. Nonetheless, the eccentricities and
inclinations appear to be coupled and to switch between modes, as was
seen in $\S$~\ref{sec:hypo}. Systems AA and AC show similar behavior
with similar amplitudes. Of the previous systems, HD~73526 is most
similar to System B (\cf Fig.~\ref{fig:104}).

\begin{deluxetable*}{ccccccccc}
\tablecaption{Initial Conditions for Selected HD~73526 Cases}
\tablecolumns{9}
\tablehead{\colhead{System} & \colhead{Body} & \colhead{$m$ (M$_{Jup}$)} & 
	\colhead{$a$ (AU)} & \colhead{$e$} & \colhead{$i$ ($^\circ$)$^f$} & 
	\colhead{$\Omega$ ($^\circ$)$^f$} & \colhead{$\omega$ ($^\circ$)} & 
	\colhead{$\mu$ ($^\circ$)}}
\startdata
AA & b & 2.89 & 0.66857 & 0.1728 & 1.37 & 182.66 & 193.96 & 96.27\\    
   & c & 2.576 & 1.0615 & 0.1126 & 1.21 & 2.44 & 147.97 & 107.05\\
AB & b & 2.881 & 0.64347 & 0.1824 & 1.27 & 262.07 & 57.63 & 95.97\\    
   & c & 2.405 & 1.0311 & 0.05134 & 1.18 & 81.74 & 8.87 & 92.51\\
AC & b & 2.87 & 0.6572 & 0.1552 & 3.37 & 342.84 & 349.04 & 75.29\\     
   & c & 2.691 & 1.0274 & 0.08992 & 2.85 & 162.79 & 324.05 & 77.07\\
\tablenotetext{f}{Measured relative to invariable plane}
\label{tab:hd73526}
\end{deluxetable*}

\subsection{The 3:1 System HD~60532}

\begin{figure*}[h] 
\begin{tabular}{cc}
\includegraphics[width=0.49\textwidth]{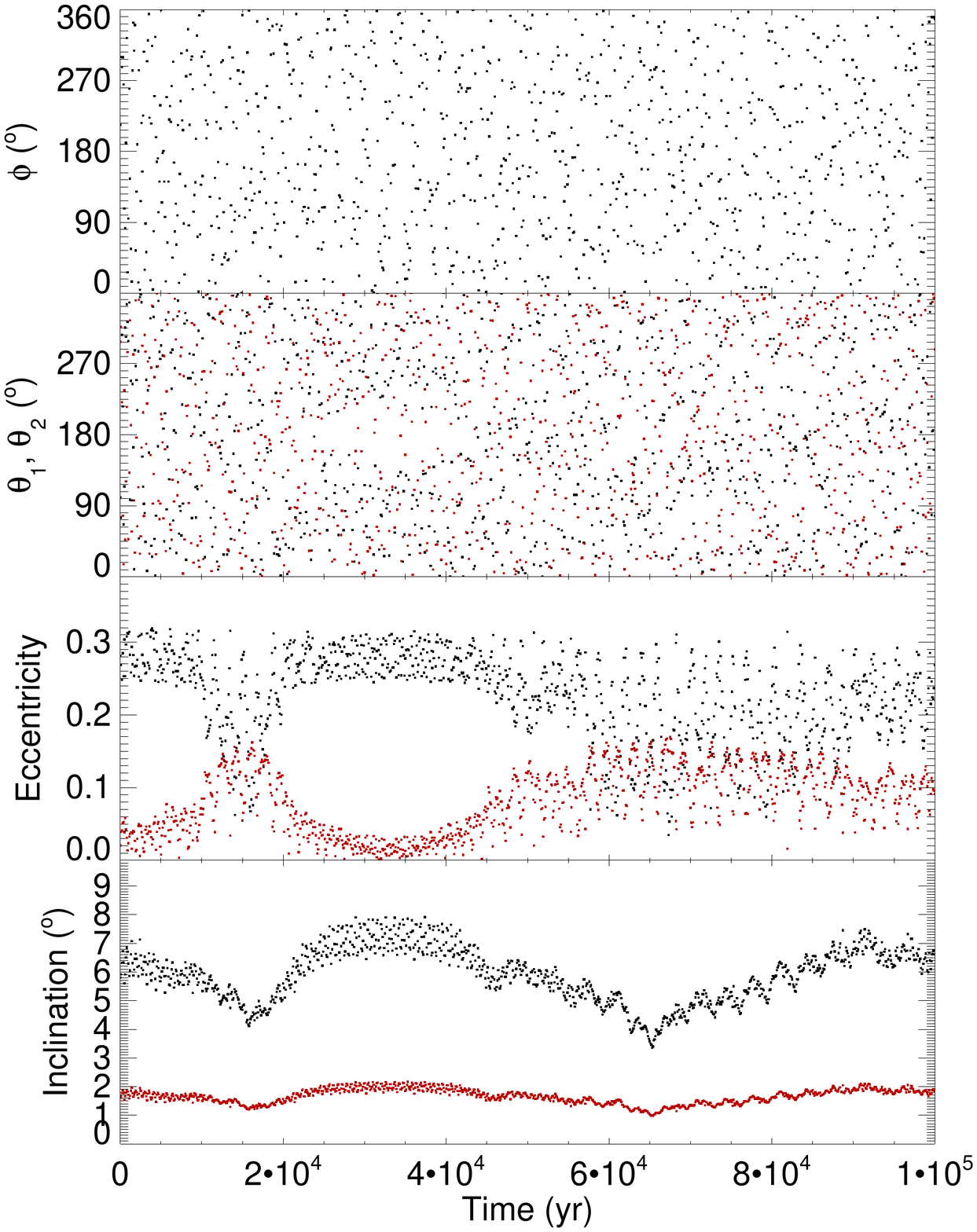} &
\includegraphics[width=0.49\textwidth]{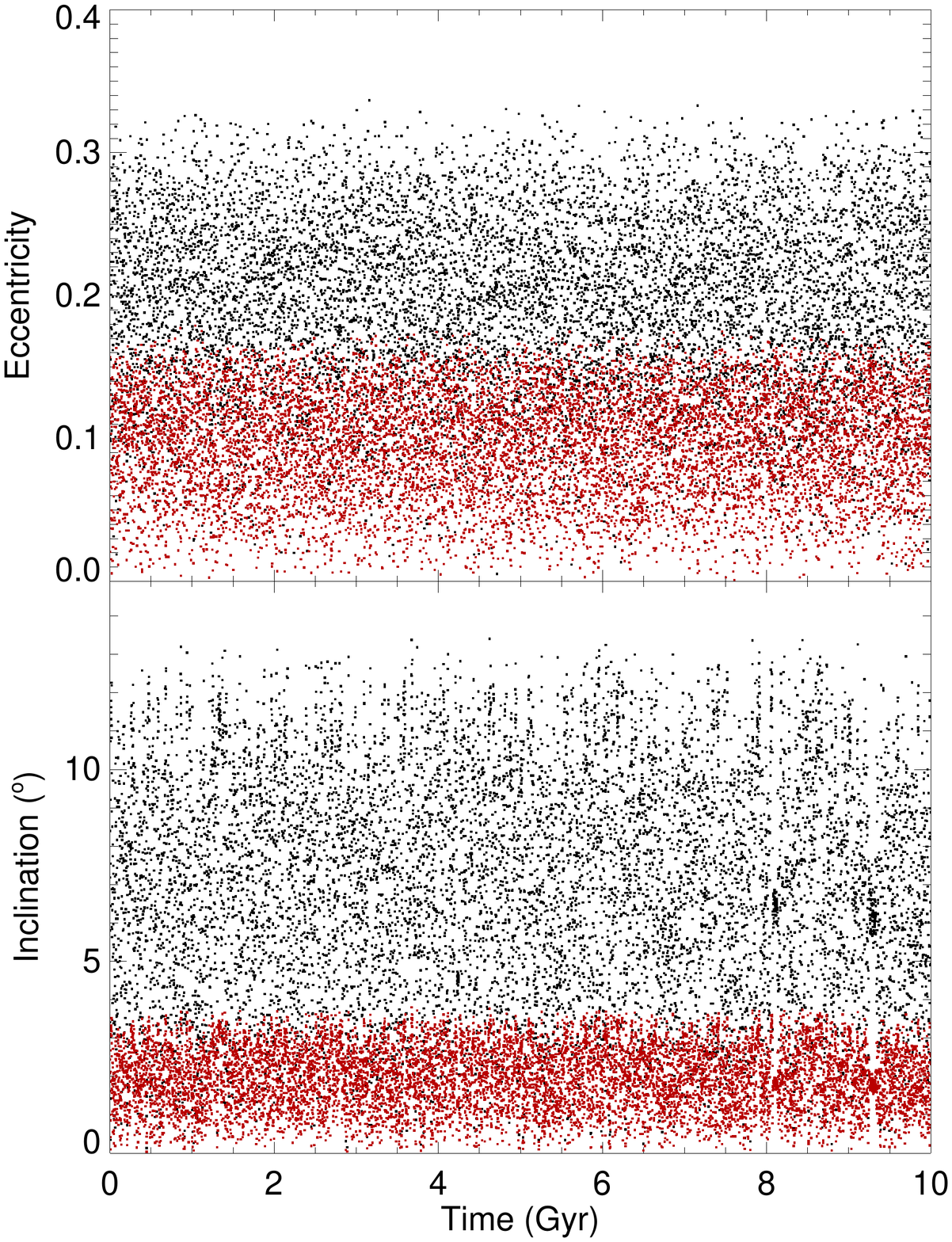}
\end{tabular}
\caption{Orbital evolution of System BC (HD~60352) in the same format as Fig.~\ref{fig:105_10Gyr}.}
\label{fig:hd60532}
\end{figure*}

In this section we consider the HD~60532 system which is in the 3:1
MMR. In Table~\ref{tab:hd60532} we list 3 cases which show chaos for 10~Gyr, and
present the evolution of case BC in Fig.~\ref{fig:hd60532}. As with
HD~75326, the $i$-resonance arguments circulate, but the $e$-arguments
switch between libration and circulation. This system appears
qualitatively similar to those in $\S$~\ref{sec:hypo}.2. Over 10~Gyr,
the system evolves chaotically, as shown in the right panels. The BA
and BB systems are qualitatively similar, but the mode-switching is
not as dramatic.

\begin{deluxetable*}{ccccccccc}
\tablecaption{Initial Conditions for Selected HD~60532 Cases}
\tablecolumns{9}
\tablehead{\colhead{System} & \colhead{Body} & \colhead{$m$ (M$_{Jup}$)} & 
	\colhead{$a$ (AU)} & \colhead{$e$} & \colhead{$i$ ($^\circ$)$^f$} & 
	\colhead{$\Omega$ ($^\circ$)$^f$} & \colhead{$\omega$ ($^\circ$)} & 
	\colhead{$\mu$ ($^\circ$)}}
\startdata
BA & & b 0.997 & 0.759 & 0.3037 & 2.56 & 267.21 & 185.73 & 162.6\\        
   & c & 2.42 & 1.5943 & 0.1578 & 0.692 & 87.1 & 252.05 & 322.57\\
BB & b & 1.053 & 0.7582 & 0.2978 & 1.403 & 253.26 & 212.28 & 277.17\\        
   & c & 2.464 & 1.5661 & 0.0354 & 0.399 & 73.39 & 250.17 & 322.34\\
BC & b & 1.062 & 0.7587 & 0.2816 & 5.806 & 293.03 & 79.61 & 209.64\\         
   & c & 2.545 & 1.572 & 0.0207 & 1.62 & 113.05 & 341.14 & 321.88\\
\tablenotetext{f}{Measured relative to invariable plane}
\label{tab:hd60532}
\end{deluxetable*}

\subsection{The 3:2 System HD~45364}

\begin{figure*}
\begin{tabular}{cc}
\includegraphics[width=0.49\textwidth]{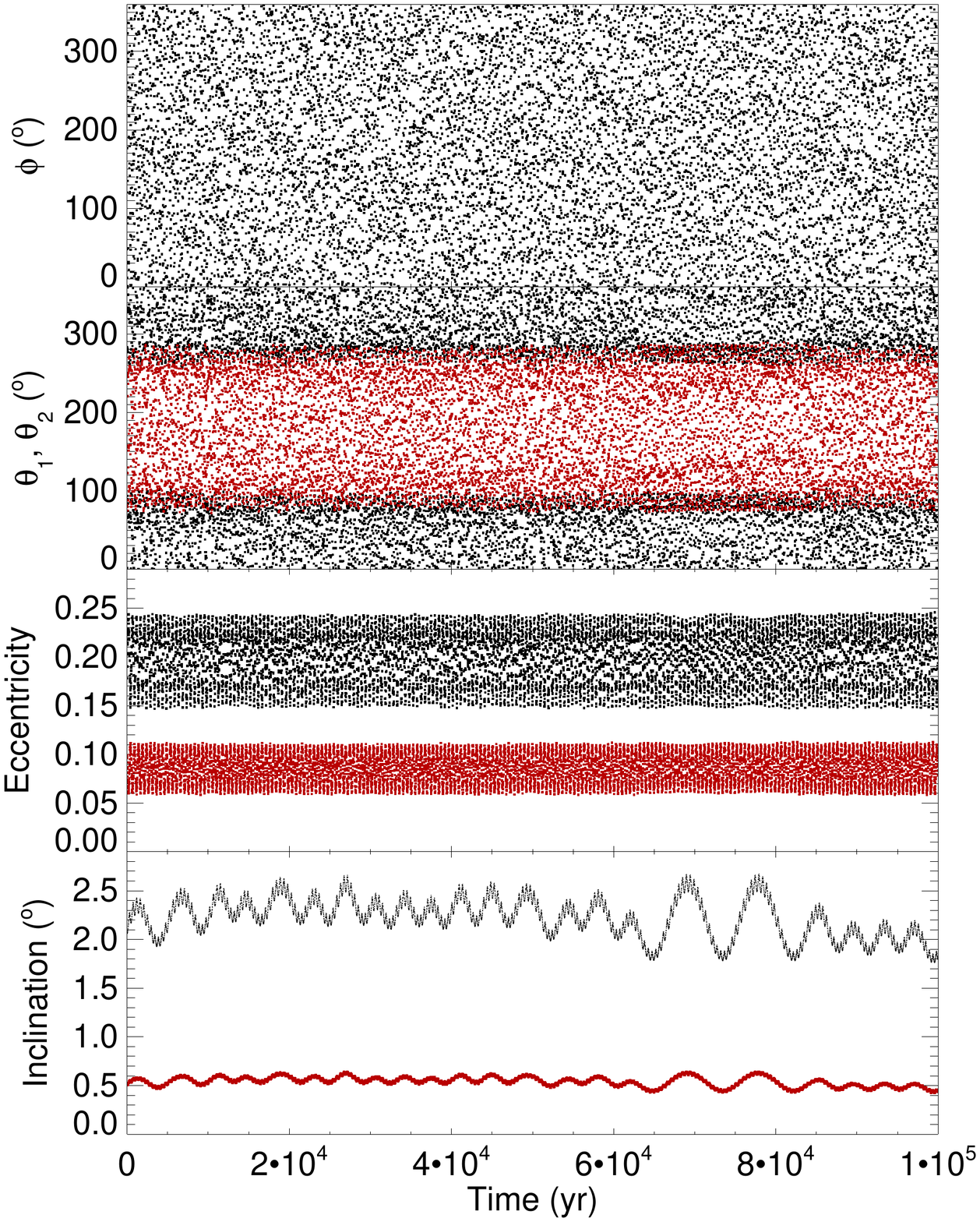} &
\includegraphics[width=0.49\textwidth]{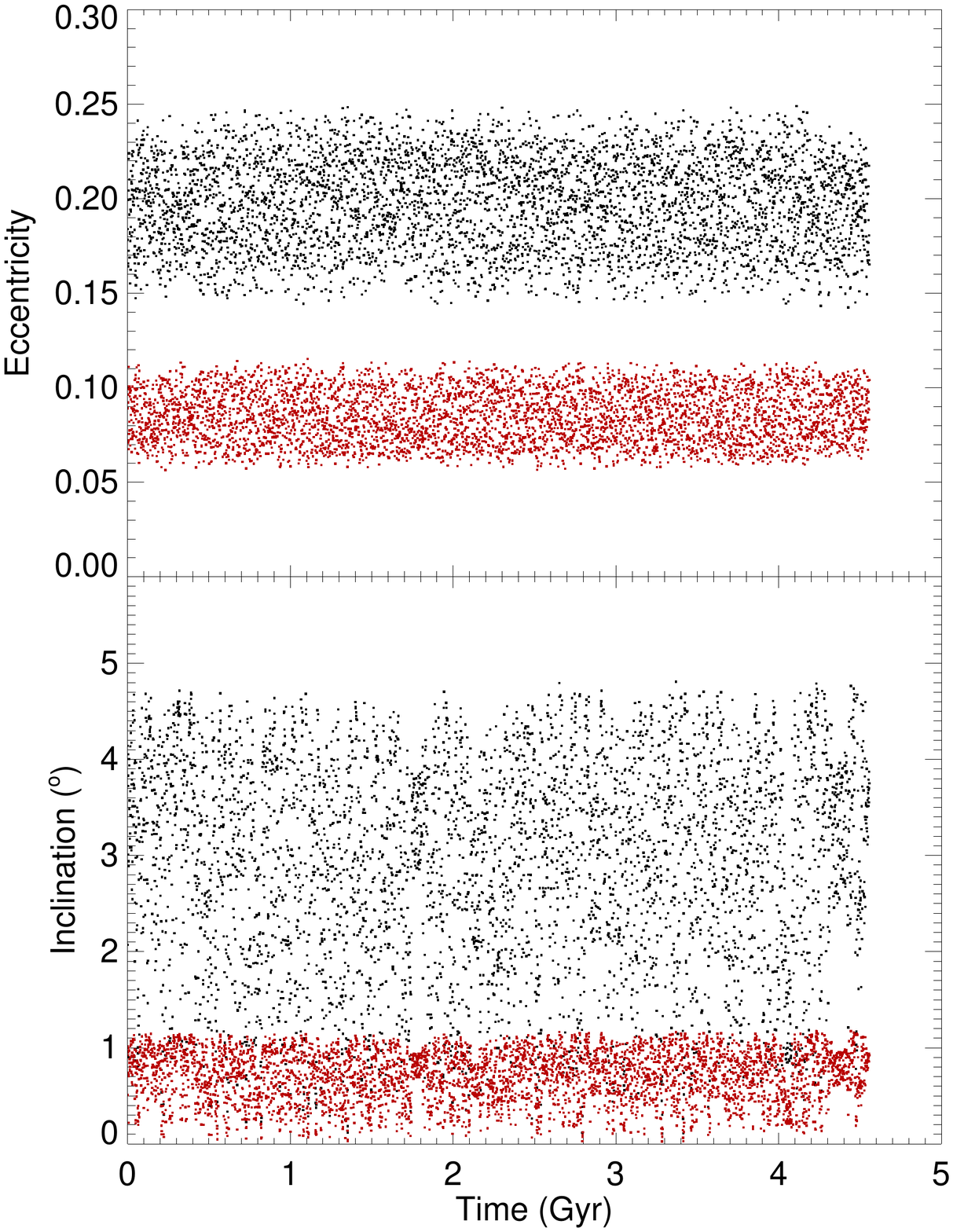}
\end{tabular}
\caption{Orbital evolution of the HD~45364 case in the same format as Fig.~\ref{fig:TS_191}.}
\label{fig:hd45364}
\end{figure*}

We found no configurations of HD~45364 that were stable for 10~Gyr, but one trial did survive for 4.557~Gyr while conserving energy to 1 part in $10^6$. Its initial conditions are in Table~\ref{tab:hd45364}, and its evolution is shown in Fig.~\ref{fig:hd45364}. 

\begin{deluxetable*}{cccccccc}
\tablecaption{Initial Conditions for the HD~45364 Case}
\tablecolumns{8}
\tablehead{\colhead{Body} & \colhead{$m$ (M$_{Jup}$)} & \colhead{$a$ (AU)} & 
	\colhead{$e$} & \colhead{$i$ ($^\circ$)$^f$} & 
	\colhead{$\Omega$ ($^\circ$)$^f$} & \colhead{$\omega$ ($^\circ$)} & 
	\colhead{$\mu$ ($^\circ$)}}
\startdata
b & 0.1874 & 0.6822 & 0.1782 & 2.07 & 359.31 & 110.52 & 199.03\\    
c & 0.6581 & 0.8969 & 0.09935 & 0.509 & 179.38 & 320.73 & 265.9\\
\tablenotetext{f}{Measured relative to invariable plane}
\label{tab:hd45364}
\end{deluxetable*}

\section{Discussion}
\label{sec:discussion}

The previous three section have demonstrated that 1) planets in an MMR
and with mutual inclinations can experience chaotic evolution of
orbital elements for Gyr while at least 1 resonant argument librates
throughout, 2) planet-planet scattering, in which one planet is
removed from a planetary system by gravitational interactions, can
leave behind two planets in the 2:1 MMR and with significant mutual
inclinations, and 3) several systems known to be in an MMR could have
mutual inclinations that induce chaotic evolution, but maintain the
resonance. In this section, we discuss the theoretical and
observational implications of these results, as well as describe the
limits of our analysis, which naturally leads to directions for future
research on this topic.

Fig.~\ref{fig:indepth} shows that conjunction can librate about
multiple centers, some of which, such as $\varpi_1$, are predicted by
classic celestial mechanics. These libration centers are derived
from low-order expansions of the ``disturbing function'' (for a review
see \cite{MurrayDermott99}). If one term dominates, there are specific
stable longitudes for conjunction, \eg $\varpi_1$ if the term
$2\lambda_2 - \lambda_1 - \varpi_1$ dominates. However, with large and
varying values of $e$ and $i$ multiple terms are important. Stable
longitudes can migrate or become unstable, allowing transitions of
libration to different kinematic modes.  This movement could be due to
the deepening of nearby minima as $e$ and/or $i$ changes. As the
orbits evolve, conjunction could gain access to another minimum,
leading to a change in the libration center.

The behavior has multiple drivers that each behave like a pendulum. In
effect, the system is analogous to a compound pendulum, \eg
Eq.~(\ref{eq:potential}). As the system evolves, the planets are able
to move into different modes of oscillation. For some modes, the
stable longitude is represented by classic longitudes like
$\varpi_1$, but others may only be derivable using higher order
theory. Our hypothesis should be testable from derivation of
high-order models of resonant behavior from the disturbing
function. Such an analysis was beyond the scope of this work, which is
just a demonstration of the amplitude and duration of the chaos, but
is clearly desirable.

This result has important implications for theoretical work on orbital
stability. Common approaches for identifying unstable orbits rely on
short integrations ($\sim 1000$ orbits) and a subsequent analysis of
the orbital evolution to count the number of frequencies in the
orbital oscillations: a larger number could indicate the system is
chaotic and unstable. Examples include the Mean Exponential Growth of
Nearby Orbits (MEGNO; \cite[\eg][]{CincottaSimo00,Gozdziewski01}) and
frequency maps \cite[\eg][]{Laskar90,LaskarCorreia09}. Our results
suggest that those methods are susceptible to labeling long-lived
systems as short-lived. The investigation of viable orbital
architectures of planets in an inclined MMR should seek configurations
that can persist for the age of the host star as described here, which
appear unstable on the short term, but are in fact long-lived. For
example, HD~202206 is likely in a 5:1 MMR, but \cite{Correia05} used a
frequency analysis to constrain the orbital parameters rather than
N-body models. Future work should explore the the veracity of these
approximate methods for the case of inclined MMRs.

At epochs of very high $e$, we expect some planets to tidally
circularize. For example, Systems D and U could not persist as shown
because tides would circularize the planet. Two scenarios are
plausible, depending on the dissipation rate in the planet. If the
dissipation is relatively weak, the resonance may be maintained such
that the resonant pair migrates inward together. Such systems should
be represented in radial velocity and transit surveys, which are
biased toward the detection of planets on relatively short orbital
periods. The three cases presented in $\S$~5 are possible examples. In
principle, the inward migration could be arrested if the tidal
dissipation forces the pair into a configuration that diminishes the
maximum eccentricity. Thus, planets found far from the host star today
could nonetheless have formed at larger distances and moved in. As the
dissipation can be episodic, it could take a long time for the pair to
migrate inward significantly, further increasing the likelihood to
detect the planets where tidal forces are expected to be
insignificant. This possibility has been discussed for non-resonant
systems \citep{Li14,DawsonChiang14}, and we find it can also occur for
MMRs. Future work should explore the role of tidal dissipation in
these systems and determine if there is any predicted or observed
signature of tidal evolution resulting from an inclined MMR.

On the other hand, if the tidal dissipation in the planet is very
large, it may break the resonance, resulting in rapid tidal
circularization and migration, leaving one close-in planet and one
more distant a planet. Although many close-in planets are singletons
\citep{Steffen12}, some are known to host more distant companions. These
companions have been suggested as perturbers that could drive eccentricities to high values through Kozai-type oscillations \citep{TakedaRasio05}, but
the possibility of resonant excitation of eccentricity has not
previously been proposed.

Note that these periods of high-$e$ can occur when $i$ has a wide
range of values. Thus, the planet's final orbital plane could be
independent of its initial orbital plane and be misaligned with its
host star's spin axis. Such spin-orbit misalignment is observed
\citep{Triaud10,Hirano14}, and numerous mechanisms have been proposed, such
as planet-planet scattering followed by tidal circularization
\citep{Chatterjee08}, tidal circularization during phases of high
eccentricity due to interactions with very distant perturbers
\citep{FabryckyTremaine07,Storch14}, and interactions with the gas
disk and a distant stellar binary companion
\citep{Batygin12}. The chaotic evolution of $e$ and $i$ in an inclined
MMR is another process to add to this list. However, given the low
occurrence rate of chaotic inclination resonances formed by
scattering, it is unlikely to be a dominant mechanism. But note that
this inference is based on only one set of scattering simulations with
specific initial conditions -- others may be more efficient and
producing these types of systems.

Inclined MMRs could impact the distribution of period ratios of
planets found via transit by making one planet's orbital plane
significantly different from the other. After applying a geometric
transit correction, \cite{SteffenHwang14} find a relative excess of
planets in the 3:2 resonance and relative deficits in 2:1 and 3:1 in
{\it Kepler} data. However, for the systems in which only 2 planets
are detected, the excess of 3:2 pairs disappears, the 2:1 is still
depleted, and the 3:1 appears to be unaffected. If additional planets
destabilize inclined MMRs, then the trend in the 3:2 MMR may be
indicative of inclined MMRs, as we would not expect systems that
evolve with large amplitude to have companions nearby. However,
caution is necessary when interpreting these data, as knowledge of the
underlying population and the role of tidal evolution is
critical. Migration during the protoplanetary disk phase often leads
to capture into resonance \citep{Snellgrove01,LeePeale02}, producing a
primordial excess of planets in MMRs. On the other hand, tidal
evolution of planets in MMRs tends to pull them to period ratios that
are not exactly commensurate
\citep{LithwickWu12,BatyginMorbi13,DelisleLaskar14}. Moreover, the
formation of inclined resonant orbits of close-in planets is unknown,
so they could be intrinsically rare. Thus, it is not obvious that
inclined MMRs are sculpting the close-in exoplanet population, but our
results suggest that they could.

Future work should identify boundaries to the long-lived but chaotic
resonances discovered here and employ a quantitative description of
the chaos.  All of our simulations of hypothetical systems began with
planetary orbital periods at exact commensurability, but this state is
not typically observed for exoplanets, see Table~\ref{tab:known} and $\S$~5. A large
suite of N-body simulations that integrate systems to 10 Gyr is
required to map out the boundaries of the chaotic and resonant
behavior. Throughout this study we have referred to systems as chaotic
as they clearly are. However, as a system moves toward regular
motion, the motion might no longer be obviously chaotic. The use of
Lyapunov exponents or other metrics would be necessary to elucidate the
true boundaries of the long-lived chaotic motion.

Identifying inclined MMRs in transit and/or RV data is possible but difficult \citep[see \eg][]{Dawson14},
but astrometric measurements are probably the best route to find their
existence. {\it HST} has successfully detected some planets for which
RV detections suggest an MMR is present \citep{Benedict02,McArthur14}
and two planets not in an MMR \citep{McArthur10}, but has not detected
two planets in an MMR. The outer planet of HD~128311 has been
detected, and the inner is potentially detectable with {\it
HST} \citep{McArthur14}, but its precarious architecture relative to
dynamical instability suggests it is in a coplanar configuration. The
most likely instrument to discover planets in an inclined MMR is {\it
GAIA} \citep{Casertano08}, as shown in Fig.~\ref{fig:astrom}. The
known systems we examined in $\S$~\ref{sec:known} are all good
candidates for {\it GAIA} astrometry, and so in the next 5 years we
should find out if any of them could be experiencing chaotic evolution.

We note that our simulations of known systems failed to reveal any in
which the $i$ arguments librate. Instead the $e$-arguments switch
between circulation and libration, which likely drives the chaos. Note
that System B evolves in a similar manner, and many other cases do as
well. Of our 3 systems formed by scattering, 2 were similar to
the known systems (Fig.~\ref{fig:scat383}). A further
exploration of known systems, either by including more or broadening
the parameter space survey, could reveal if this trend is real. If
known systems are only consistent with $i$-argument circulation, it
could provide important clues to the formation and frequency of
exoplanets in inclined MMRs.

Throughout this study we have described systems that survive for
10~Gyr as ``stable.'' As most of our host stars are solar analogs,
this usage is reasonable as the systems survive for the main sequence
lifetimes of the host stars. However, in some cases we find
destabilization can occur on $>1$~Gyr timescales. Hence, these
systems are not ``stable'' in the sense that they could survive
indefinitely. Late-term destabilization of systems in inclined MMRs
may explain the observation that the host stars of planets in the 2:1
MMR tend to be younger than other host
stars \citep{KoriskiZucker11}. If this observation is validated, it
would support the hypothesis that some MMRs evolve chaotically and
disrupt on long timescales.
 
In the previous sections we did not consider the role of additional
planets, which could significantly shrink the longevity of a
chaotically evolving system. It is clear that for system in which
$e_1~\sim~1$ that interior planets are forbidden. However, exterior
planets could exist provided they are distant and/or small. On the
other hand, some systems show low amplitude chaotic variations and may
therefore be robust to perturbations from a third planet. We do not
map out the role of a third companion in the stability of our systems, but
future work should explore how they modify the results presented here.

Most of our simulations begin with an Earth-mass planet at 1~AU from a
solar-mass star, and would be considered potentially
habitable \citep{Kasting93,Kopparapu13}. If habitable, these worlds
would be markedly different from the Earth, with unpredictable
climates on geologic timescales. For planets in which the eccentricity
grows large, the planet could occasionally enter a runaway greenhouse (incident
radiation flux scales as $(1-e^2)^{-1/2}$) rendering the planet
uninhabitable. Large inclination fluctuations can also lead to large
variations in obliquity, which is a major driver of climate
evolution \citep[\eg][]{WilliamsKasting97,Spiegel09}. While extremely fast and large
variations of obliquity could be detrimental to habitability, at the
outer edge of the habitable zone, these variations can suppress ice
sheet growth and, in principle, increase a planet's
habitability \citep{Armstrong14}. For planets that occasionally reach very high eccentricity, tidal dissipation could ultimately pull the planet out of the habitable zone \citep{Barnes08}. Should any potentially habitable
planets be found in an MMR, it will be imperative to understand the
orbital evolution, and its connection to climate, prior to investing
spectroscopic observations of its atmosphere to search for
biosignatures \citep[see \eg][]{Deming09,KalteneggerTraub09,Misra14}.

\section{Conclusions}
\label{sec:concl}

We have simulated the orbital evolution of exoplanets in mean motion
resonances and inclinations and found the orbits can evolve
chaotically for at least 10~Gyr. We hypothesize that these systems
behave like compound pendula, which are naturally chaotic systems that
can switch between modes of oscillation, as seen in our simulations
(see Fig.~\ref{fig:indepth}). We find this chaotic motion over a range
of mass ratios and for the 2:1, 3:2 and 3:1 resonance.  We also tested
different N-body codes using different integration schemes, and
conclude the results are robust. Inclined MMRs can be produced by
planet-planet scattering and the resultant systems are qualitatively
similar to our simulations of known systems in an MMR.

These results have numerous implications for both theory and observations:
\begin{itemize}
\item{Approximate methods to estimate stability with short integrations may be unreliable near MMRs.}
\item{Close-in planets may arrive at their current orbits due to eccentricity excitation by inclined MMRs followed by rapid tidal circularization.}
\item{Some short-period planets with orbital planes misaligned with the stellar spin axis may be produced by systems initially in inclined MMRs.}
\item{MMR pairs may be episodically migrating inward due to weak dissipation occurring during epochs of very large eccentricity.}
\item{Systems in an MMR may be systematically younger than other multiplanet systems due to the destabilization of older MMR systems.}
\item{The distribution of period ratios of adjacent planets detected via transit may be skewed by inclined MMRs.}
\item{Potentially habitable planets may be severely impacted by the orbital architecture of the system.}
\end{itemize}

Although no systems are currently known to demonstrate the behavior we have outlined here, the {\it GAIA} space telescope has the power to detect hundreds of giant exoplanets in inclined MMRs.

\medskip

This work was supported by NASA's Virtual Planetary Laboratory under Cooperative Agreement No. NNA13AA93A and NSF grant AST-1108882.

\bibliography{InclRes}

\end{document}